\documentclass[twocolumn,showpacs,showkeys, amsmath,amssymb,footinbib]{revtex4}
\usepackage{graphicx}
\usepackage{dcolumn}
\usepackage{bm}
\usepackage{epsf}

\newcommand{\be}{\begin{equation}}
\newcommand{\ee}{\end{equation}}
\newcommand{\bea}{\begin{eqnarray}}
\newcommand{\eea}{\end{eqnarray}}
\newcommand{\ba}{\begin{array}{ccc}}
\newcommand{\ea}{\end{array}}
\newcommand{\nn}{\nonumber}

\newcommand{\eq}[1]{eq.~(\ref{#1})}
\newcommand{\eqs}[2]{eqs.(\ref{#1}, \ref{#2})}

\newcommand{\Eq}[1]{Eq.~(\ref{#1})}
\newcommand{\Eqs}[2]{Eqs.(\ref{#1}, \ref{#2})}

\newcommand{\fig}[1]{Fig.\ref{#1}}
\newcommand{\ur}[1]{(\ref{#1})}
\newcommand{\urs}[2]{(\ref{#1},\ref{#2})}
\newcommand{\half}{{\textstyle{\frac{1}{2}}}}

\newcommand{\skp}{\sum\hspace{-.45 cm}\int}
\newcommand{\skpm}{\sum\hspace{-.45 cm}\int^{M}}

\def\log{\textnormal{log}}
\def\det{\textnormal{det}}
\def\exp{\textnormal{exp}}

\def\ge{\gamma_{{\rm E}}}
\def\Tr{ {\rm Tr} }
\def\Sp{{\rm Sp}}

\def\cos{{\rm cos}}

\def\sin{{\rm sin}}
\def\A{{\cal A}}

\def\appb#1#2#3{Acta Phys. Pol. B {\bf #1}, #2 (#3)}

\def\ibid#1#2#3{{\it ibid.} {\bf #1}, #2 (#3)}

\def\jetp#1#2#3{JETP Lett. {\bf #1}, #2 (#3)}

\def\npb#1#2#3{Nucl. Phys. B {\bf #1}, #2 (#3)}

\def\plb#1#2#3{Phys. Lett. B {\bf #1}, #2 (#3)}
\def\pr#1#2#3{Phys.Rev.{\bf #1}, #2 (#3)}

\def\prd#1#2#3{Phys. Rev. D {\bf #1}, #2 (#3)}
\def\prl#1#2#3{Phys. Rev. Lett. {\bf #1}, #2 (#3)}

\def\pzetf#1#2#3{Pisma Zh. Eksp. Teor. Fiz. {\bf #1}, #2 (#3)}

\def\rmp#1#2#3{Rev. Mod. Phys.{\bf #1}, #2 (#3)}
\def\sjnp#1#2#3{Sov. J. Nucl. Phys.  {\bf #1}, #2 (#3)}
\def\yadf#1#2#3{Yad. Fiz. {\bf #1}, #2 (#3)}
\def\zpc#1#2#3{Z. Phys. C {\bf #1}, #2 (#3)}

\begin{document}

\title{Gauge invariant effective action for the Polyakov line 
in the SU(N) Yang--Mills theory at high temperatures}
\author{Dmitri Diakonov$^{a,b}$}
\email{diakonov@nordita.dk}
\author{Michaela Oswald$^{c}$}
\email{oswald@alf.nbi.dk}
\affiliation{
$^a$NORDITA,  Blegdamsvej 17, DK-2100 Copenhagen,
Denmark\\
$^b$St. Petersburg NPI, Gatchina, 188 300, St. Petersburg, Russia\\
$^c$NBI, Blegdamsvej 17, DK-2100 Copenhagen, Denmark
}

\date{September 14, 2004}

\begin{abstract}
We integrate out fast varying quantum fluctuations around static $A_4$ and
$A_i$ fields for the $SU(N)$ gauge group. By assuming that the gluon
fields are slowly varying but allowing for an arbitrary amplitude of $A_4$ we obtain 
two variants of the effective high-temperature theory for the Polyakov line.
One is the effective action for the gauge-invariant eigenvalues of the Polyakov
line, and it is explicitly $Z(N)$ symmetric. 
The other is the effective action for the Polyakov line itself as an 
element of the $SU(N)$. In this case the theory necessarily includes the spatial 
components $A_i$ to ensure its gauge invariance under spatial gauge transformations. 
We derive the 1-loop effective action in the `electric' and `magnetic' sectors, 
summing up all powers of $A_4$.
\end{abstract}


\maketitle

\section{Introduction}
Finite-temperature quantum chromodynamics (QCD) is an intensely studied
field. At finite temperature gluons obey periodic boundary conditions 
in imaginary time, leading to the quantized Matsubara frequencies 
$\omega_k=2\pi kT$ for gluons. At very high temperatures nonzero Matsubara frequencies 
decouple from a theory as they become infinitely heavy. This is called 
dimensional reduction \cite{dimred} since the decoupled heavy modes are  
time-dependent ones. Neglecting all modes except the static zero Matsubara 
frequencies leaves a $3{\rm D}$ theory
\bea
-\frac{1}{4g^2}\int\!d^4 x\,F_{\mu\nu}^2 \to 
-\frac{1}{4g^2 T}\int\!d^3x\left[F_{ij}^2+2(D_i^{ab}A_4^b)^2\right],
\label{treelevel}
\eea
which only contains the static gluonic modes with
the coupling constant $g_{\it 3}^2 = g^2 T$. The long-range forces 
mediated by the static gluons lead to the IR divergencies, 
because in strict perturbation theory the gluons are massless.
Owing to the chromomagnetic sector, the high-temperature perturbation 
theory explodes already at a few-loop level \cite{P,L,GPY} and 
is hence only applicable at academically high temperatures \cite{breakdown}.
The region of intermediate temperatures is of much bigger interest.
For example, the deconfinement phase transition in the pure Yang--Mills theory
is believed to take place in this region. 

In the pure Yang-Mills theory the center symmetry plays a crucial part in the
description of the confinement-deconfinement phase transition
\cite{P,S,'tH,SY}. The order parameter for the latter is the average 
of the trace of the Polyakov line:
\bea\label{Polyakov_line}
L(x)={\cal P}\,\exp\left(i\int_0^{1/T}\!dx_4\,A_4\right).
\eea
The order parameter $<\!\Tr L\!>$ is zero in the
confined phase below the critical temperature and assumes, after a proper 
regularization, a non-zero value in
the deconfined phase above the critical temperature. For a recent
confirmation in lattice simulations see \cite{Karsch, DHLOP}. 
The Polyakov line is not invariant under gauge transformations 
belonging to the gauge group center. One hence concludes that 
if $<\!\Tr L\!>=0$ then the $Z(N)$ symmetry is
manifest. This situation describes confinement. If for any reason 
$<\!\Tr L\!>\neq 0$ then the symmetry must have been broken. 
This corresponds to the deconfined phase.

At high but not infinitely high temperatures the tree-level action
\ur{treelevel} is not sufficient to study gluon fluctuations. 
When one includes quantum corrections all the non-zero
Matsubara modes of the gluons show up in the loops.
The 1-loop \cite{GPY, Weiss} and 2-loop \cite{2loop} potential
energies as functions of $A_4$ are known. They are periodic functions
of the eigenvalues of $A_4$ in the adjoint
representation with period $2\pi T$. This reflects the symmetry of the
$Z(N)$ vacua. The curvature of the potential at its minima gives the
leading order Debye mass for `electric' gluons. The zero energy
minima of the potential are at quantized values of $A_4$ corresponding
to the center-of-group values of the Polyakov line, where $L\in Z(N)$. 
At high temperatures the system oscillates around one of these minima.
At lower temperatures, however, the fluctuations around the minimum
increase and eventually the system undergoes a phase transition to 
$<\!\Tr L\!>=0$. At the same time, one expects that near the phase 
transition point the fluctuations are long-range. To study those fluctuations, 
one needs an effective low-momenta theory which, however, does not 
assume that the $A_4$ component is small. 

The nonzero temperatures explicitly break the $4{\rm D}$ Euclidean symmetry 
of the theory down to the $3{\rm D}$ Euclidean symmetry, so that the spatial 
$A_i$ and time $A_4$ components of the Yang--Mills field play different 
roles and should be treated differently. One can always choose a gauge 
where $A_4$ is time-independent. Taking $A_4(x)$ to be static is not 
a restriction of any kind on the fields but merely a convenient gauge 
choice, and we shall imply this gauge throughout the paper. It is also 
a possible gauge choice at $T=0$ but in that limiting case it is unnatural
as one usually wishes to preserve the $4{\rm D}$ symmetry. In this gauge, the 
Polyakov line \ur{Polyakov_line} is not a path-ordered but a simple
exponent of $A_4(x)$. It rotates under $x$-dependent gauge transformations.
Therefore, an effective action for the Polyakov line, which should be invariant 
under spatial gauge transformations, cannot depend on the gradients
of $L(x)$ alone, but rather on the {\em covariant} derivatives of $L(x)$,
involving the spatial components of the background field $A_i$. The aim
of this work is to compute the effective action expanding it in covariant
derivatives but {\em keeping all powers of} $A_4$. It is an extension 
of the previous work \cite{DO2G,MO,DO2F} to higher gauge groups. 
We use the background field method for the gluons and evaluate the 1-loop
action through the functional determinant formalism. We integrate out 
fast varying quantum fluctuations about the static background $(A_i,A_4)$
by making an expansion in spatial covariant derivatives. This method was originally 
developed in \cite{DPY} for zero temperature QCD. 
We compute terms quadratic in covariant derivatives, that is the `electric' $E^2$ 
terms where $E_i^a=D_i^{ab}A_4^b$, and the quartic `magnetic' $B_{\parallel}^2$ terms where
$B_i^a=\half\epsilon_{ijk}(\partial_iA_j^a-\partial_jA_i^a+f^{abc}A_i^bA_j^c)$.
While we consider a general electric field, we restrict ourselves to a magnetic field parallel to $A_4$, $B_i=B_{i\parallel}$.
The structure of the $SU(3)$ effective action turns out to be much 
richer than in the $SU(2)$ case, as there are far more invariants 
which one can build out of $A_4$ and the electric and the magnetic fields. 

As a byproduct of our study, we obtain the effective action for the
{\em eigenvalues} of the Polyakov line. Contrary to the Polyakov line itself
which is an element of the $SU(N)$ group and rotates under spatial gauge
transformations, its eigenvalues are gauge invariant. Therefore, the 
derivative expansion of the effective action for the eigenvalues goes in
ordinary rather than covariant derivatives. Such an action was first obtained
in ref. \cite{BGK-AP} for the $SU(2)$ gauge group and reaffirmed in
ref. \cite{DO2G}; here we extend it first to the more intricate case of 
the $SU(3)$ gauge group and then to the general $SU(N)$. 
The resulting action is explicitly $Z(N)$ symmetric.  

We expect that our results are suitable to study the
correlation functions of the Polyakov line not too far from the transition
point where it experiences fluctuations that are large in amplitude 
but presumably mainly long ranged.

\section{Covariant derivative expansion of 1-loop quantum action}

In \cite{DO2G} we obtained the 1-loop effective action for the pure Yang--Mills
$SU(2)$ theory. We started with the Yang--Mills partition function and 
decomposed the gluon fields into background fields and quantum
fluctuations around them. For a 1-loop approximation we expanded the action to
quadratic order in the gluonic quantum fluctuations. The resulting bilinear
form for the quantum fluctuations is degenerate, so we chose a 
background Lorenz gauge for them. This in turn gives rise to the Faddeev-Popov ghost
determinant in the background field. The 1-loop effective action was 
obtained by integrating over the ghost fields as well as over the gluon 
quantum fluctuations. It is expressed in terms of functional determinants 
which is an economic and aesthetic method of getting gauge invariant results 
\footnote{Our convention for the sign of the action is such that the partition 
function ${\cal Z} = \exp(+S)$.}:
\be\label{S1l}
S_{\rm{ 1-loop}} = \log\, \left(\det{W}\right)^{-1/2} + \log\,\det \left(-D^2\right)\;.
\ee
The first contribution here comes from integrating out the gluon fields,
while the second comes from the ghost fields. 
Since the gluons transform under the adjoint representation of the color
group the operators $D^2, W_{\mu\nu}$ consist of matrices in the adjoint
representation. The latter is given by
\be\label{W}
W_{\mu\nu}^{ab} = -[ D^2(A)]^{ab} \delta_{\mu\nu} - 2 f^{acb}F_{\mu\nu}^c(A) \;,
\ee
where
\be\label{D}
D_{\mu}^{ab} = \partial_{\mu}\delta^{ab} + f^{acb}A_{\mu}^c 
\ee
is the covariant derivative in the adjoint
representation. All gluon fields $A_\mu$ are background fields. 
\Eq{S1l} is independent of the gauge group and hence valid for any
$SU(N)$. In addition, it is invariant under general gauge transformations of the
$A_\mu$ fields. We can use this property to make the $A_4$ component
static. Then, there still remains a freedom to make time-independent gauge
transformations,
\bea\nonumber
A_4(x) &\to & U(x)\,A_4(x)\,U^\dag(x),\\
A_i(x) &\to & U(x)\,A_i(x)\,U^\dag(x)+i\,U(x)\,\partial_i\,U^\dag(x)\;,
\label{GT}\eea
which can be used to bring $A_4$ to a diagonal form in the fundamental representation.
The general $SU(N)$ case is considered in section XI,B. 
For the $SU(3)$ gauge group we can write ($\lambda^a$ are 8 Gell-Mann matrices, see
the Appendix) 
\be\label{diagform}
A_4=A_4^3(x)\frac{\lambda^3}{2}+A_4^8(x)\frac{\lambda^8}{2}.
\ee
This gauge fixing leaves 
$A_i$ an arbitrary $SU(3)$ matrix, generally speaking both space and time dependent. 

After fixing the gauge such that $A_4$ is diagonal and static, one is still left 
with a freedom of gauge transformations of a special type. For the $SU(2)$ group
they were pointed out in ref. \cite{DO2G}, section 10; in the $SU(3)$ case the
remaining gauge transformations are of the form
\bea\label{discreteGT}
A_\mu &\to & S A_\mu S^\dagger+iS\partial_\mu S^\dagger,\\\nonumber
S(x,t)\!\!\!&\!=\!&\!\!\!\exp\!\left\{\!i\frac{\lambda^3}{2}\left[\alpha_3(x)\!\!+\!\!2\pi tTn_3\right]
\!\!+\!\!i\frac{\lambda^8}{2}\left[\alpha_8(x)\!\!+\!\!2\pi tTn_8\right]\right\},
\eea    
where $n_3$ and $n_8\sqrt{3}$ are both even or both odd integers. One cannot take rotations about 
axes other than the $3^{\rm rd}$ and $8^{\rm th}$ ones because it will make $A_4$ non-diagonal, 
and one cannot take the time dependence other than linear because that would make $A_4$ 
time dependent. The space-dependent functions $\alpha_{3,8}$ may be arbitrary.
The fact that $n_3$ and $n_8\sqrt{3}$ must be both even or both odd integers 
follows from the requirement that the gauge transformation in the adjoint representation, 
given by the $8\times 8$ matrix $O^{ab}=2\Tr(St^aS^\dagger t^b)$, must be periodic in time. 
In components, the transformation \ur{discreteGT} reads:
\bea
\label{A438}
A_4^{3,8}&\to& A_4^{3,8}+2\pi T\, n_{3,8},\\
\label{Ai38}
A_i^{3,8}&\to & A_i^{3,8}+\partial_i\alpha_{3,8}, \\
\label{Ai1}
A_i^1&\to & \cos\, \beta_3\,A_i^1+\sin\,\beta_3\,A_i^2,\\
\label{Ai2}
A_i^2&\to & -\sin\, \beta_3\,A_i^1+\cos\,\beta_3\,A_i^2,\\
\label{Ai4}
A_i^4&\to & \cos\frac{\beta_8+\beta_3}{2}\,A_i^4
+\sin\frac{\beta_8+\beta_3}{2}\,A_i^5,\\
\label{Ai5}
A_i^5&\to & -\sin\frac{\beta_8+\beta_3}{2}\,A_i^4
+\cos\frac{\beta_8+\beta_3}{2}\,A_i^5,\\
\label{Ai6}
A_i^6&\to & \cos\frac{\beta_8-\beta_3}{2}\,A_i^6
+\sin\frac{\beta_8-\beta_3}{2}\,A_i^7,\\
\label{Ai7}
A_i^7&\to & -\sin\frac{\beta_8-\beta_3}{2}\,A_i^6
+\cos\frac{\beta_8-\beta_3}{2}\,A_i^7,\\
\label{beta}
\beta_{3,8}&=& 2\pi t Tn_{3,8}+\alpha_{3,8}.
\eea
The Polyakov line is transformed by a diagonal matrix
\be
{\rm diag}\left(e^{\pi i (n_3+ \frac{n_8}{\sqrt{3}})},e^{-\pi i(n_3-\frac{n_8}{\sqrt{3}})},
e^{-\frac{2\pi i}{\sqrt{3}}n_8}\right),
\label{transfPol}\ee
which for $n_3$ and $n_8\sqrt{3}$ both even or both odd becomes an element of the group
center,
${\rm diag}\left(e^{\frac{2\pi i k}{3}},e^{\frac{2\pi i k}{3}},e^{\frac{2\pi i k}{3}}\right)$,
$k=0,1,2$. 

Thus we see that the $Z(N)$ symmetry of the effective action is a consequence
of the symmetry under gauge transformations of a special type, \eq{discreteGT}. 
It should be stressed, however, that the same gauge transformations make the 
spatial components of the background field $A_i$ time dependent, 
even if one starts from purely static $A_i$'s, with their Matsubara frequencies 
directly related to $n_{3,8}$ according to eqs. (\ref{Ai1}-\ref{Ai7}). 
Consequently, if one wishes to see the $Z(N)$ symmetry 
of the effective action \ur{S1l} explicitly, one has to sum up all powers in time derivatives
of $A_i$. This seems to be a formidable problem which we do not attempt
to solve here. We expand the effective action \ur{S1l} to the second order
in the electric field and hence to the second order in $\dot A_i$ only. 
Therefore, our effective action will not be explicitly $Z(N)$ symmetric. 
To be more precise, part of the invariants we calculate will be explicitly
$Z(N)$ symmetric, namely those which remain invariant under the transformations
(\ref{Ai1}-\ref{Ai7}), but some of the invariants will be not. The effective 
action we are computing is for the static $A_i$ fields and is invariant only 
under static gauge transformations \ur{GT}. 
 
With our background-field technique, we are also able to solve simultaneously 
another physical problem. Namely, one may be interested in the effective quantum 
action for the {\em eigenvalues} of the Polyakov line. Contrary to $A_4$ and to the
Polyakov line as a unitary matrix, its eigenvalues {\em are invariant} under spatial gauge
transformations. Therefore, the effective action for the eigenvalues can be expanded
in ordinary `short' rather than covariant derivatives, and the background
$A_i$ field may then be set to zero. In this setting, the $A_i$ fields (as well as
the rapidly changing components of the $A_4$ fields) are understood and treated 
as quantum fluctuations over which one integrates. The resulting effective action
for the eigenvalues of the Polyakov line, including its spatial derivatives
to the second order, explicitly obeys the $Z(N)$ symmetry. It has been first
found for the $SU(2)$ gauge group in ref. \cite{BGK-AP} and reaffirmed in ref. \cite{DO2G};
here we extend this result to the $SU(3)$ case in section XI,A and to the general $SU(N)$
gauge group in section XI,B.  

The functional determinants in \eq{S1l} are UV divergent which
reflects the divergence of the coupling constant. Since QCD is a renormalizable 
theory the divergence can be absorbed in the definition of the tree level 
coupling constant. To do so, we have to properly
normalize and regularize the functional determinants. The former is obtained
by normalizing to the free zero gluon contribution, for the latter we introduce
a Pauli--Villars cutoff $M$ in momentum space. For the gluon determinant this
results in the ``quadrupole formula''
\bea\label{Dnr}
\lefteqn{\det (-D^2)_{\rm{r, n}}
\equiv \frac{\det (-D_{\mu}^2)}{\det (-\partial_{\mu}^2)}\,\frac{\det
 (-\partial_{\mu}^2 + M^2)}{\det (-D_{\mu}^2 + M^2)}} \\ \nonumber
&&= \exp\left\{
-\int_0^{\infty}\frac{ds}{s}\;  \Sp \left[\left( 1 - e^{-s M^2}\right) \,
\left( e^{s D_{\mu}^2} - e^{s \partial_{\mu}^2} \right)\right]\right\}\;,
\eea
where the rewriting in the last line is due to a trick originally introduced
by Schwinger \cite{Schwinger}. The functional trace Sp can be taken by
inserting any complete basis. We choose the plane wave basis:
\bea\label{planewave}
{\rm Sp}\, e^{-s K} &\!\!=\!\!& \Tr\!\!\int\!\! d^4 x 
\lim_{y\to x}\int\frac{d^4p}{(2 \pi)^4} 
e^{-ip\cdot y}e^{-s K} e^{ip\cdot x}\\
\nonumber 
&\!\!=\!\!& \Tr\!\!\int \!\!d^4 x \int\frac{d^4 p}{(2 \pi)^4} e^{-s K (\partial_{\alpha}
\rightarrow \partial_{\alpha}+ i p_{\alpha})}\bf{1}\,.
\eea
Here Tr is the remaining trace over color and Lorentz indices. 
The $\bf{1}$ at the end is meant to emphasize that the shifted operator acts on unity,
so that for example any term that has a $\partial_{\alpha}$ in 
the exponent and is brought all the way to the right, will vanish. 
According to ({\ref{planewave}) we now have
\bea\label{Dpw}
\lefteqn{\log\,\det (-D^2)_{\rm{r,n}}
=}\\ \nonumber
&& - \int d^3 x\sum_{k=-\infty}^{\infty}\int\frac{d^3 p}{(2
\pi)^3}\int_0^{\infty}\frac{ds}{s} \left( 1 - e^{-s M^2}\right)e^{-s
  p^2}\\ \nonumber
&& \times \,\Tr\,\left\{\exp\left[ s {\cal A}^2 + s D_i^2 
+ 2isp_iD_i\right]-\exp\left[-s {\omega_k}^2\right]\right\}\;,
\eea
where we introduced the adjoint matrix
\bea\label{A}
{\cal A} = f^{acb}A_4^c + i\omega_k \delta^{ab}.
\eea
Similarly to \eq{Dnr} we have to normalize and regularize the gluon
functional determinant. Again after an insertion of a plane wave basis we
obtain
\bea\label{Wpw}
\lefteqn{\log\,(\det W)^{-1/2}_{{\rm r,n}} =}\\ 
\nonumber
&&
\frac{1}{2}\int\,d^3 x\,\sum_{k=-\infty}^{\infty}\int\frac{d^3 p}{(2 \pi)^3}\,
\int_0^{\infty}\frac{ds}{s}\left( 1 - e^{-s M^2}\right)e^{-s
  p^2}\\ 
\nonumber
&&\times
\Tr\,\left\{\exp\left[(s \A^2 + s D_i^2 + 2isp_iD_i)^{ab}
    \delta_{\mu\nu} +
2sf^{acb}F_{\mu\nu}^c\right]\right.\\ \nonumber && \left.\qquad
-  \exp\left[-s {\omega_k}^2\right]\right\}\;.
\eea
The regularized and normalized effective 1-loop action is hence given by the
sum of \eq{Dpw} and \eq{Wpw}. Expressed in this way one can now expand in
powers of spatial covariant derivatives $D_i$. This introduces a potential
energy and a kinetic energy in terms of color-electric and color-magnetic fields, 
since the latter are identified as
\bea\label{E-B}
i\left[D_i, {\cal A}\right] &=& i\left[D_i,D_4\right] = F_{i4} = 
E_i\quad{\rm and}\quad \\ \nonumber
\half\epsilon_{ijk}F_{jk} &=& \frac{i}{2}\left[D_j,D_k\right] = B_i.
\eea
In \cite{DO2G} we studied the effective action for the gauge group $SU(2)$. We
expanded to quadratic order in the electric and magnetic fields, but did not
retain mixing terms between electric and magnetic field or derivative terms
of the electric field, which all exist at that order. The gauge invariant
structures that we obtained were $A_4^a A_4^a$, $E_i^a E_i^a$, $B_i^a B_i^a$,
$(E_i^a A_4^a)^2$ and $(B_i^a A_4^a)^2$. For $SU(3)$, which is the case that we
are considering here, one expects and finds more invariants.

\section{The Polyakov line}
The order parameter for the transition from a confined to a deconfined phase
in a pure Yang--Mills theory is the average of the trace of the Polyakov line,
\be\label{pline}
L(x) = {\cal P}\,\exp\left\{i \int_0^{\beta=\frac{1}{T}} dx_4 A_4(x_4, x)\right\}.
\ee
As usual ${\cal P}$ denotes path ordering. For our explicit calculations we
shall use the freedom to rotate the $A_4$ field to a diagonal form and consider
it to be static. In this gauge, $L(x)$ is a diagonal matrix $\exp(iA_4/T)$. For an
$SU(N)$ there are $(N-1)$ diagonal generators. In $SU(2)$ this is the third Pauli
matrix $\tau_3$, in $SU(3)$ one has the $\lambda^3$ and $\lambda^8$ Gell-Mann
matrices (the general $SU(N)$ case will be considered in section XIB). 
Writing $A_4$ in the basis of these diagonal generators automatically ensures 
that it is traceless. 

In particular we choose the following parametrization ($t^a=\lambda^a/2$, see
Appendix)
\be\label{a4diag}
A_4 = A_4^3 t^3 + A_4^8 t^8.
\ee
From this we can then construct $A_4$ in the adjoint representation as
$A_4^{\rm adj} =A_4^{ab} = i f^{acb}A_4^{c}$. 
We find that it has the following non-zero eigenvalues
\bea\label{a4ev}
\pm\phi_1 &=&\pm A_4^3,\\ \nn
\pm\phi_2 &=& \pm\frac{A_4^3 + A_4^8\sqrt{3}}{2}, \\ 
\nn
\pm\phi_3 &=& \pm\frac{A_4^3 - A_4^8\sqrt{3}}{2}.
\eea
Introducing the rescaled variables $a_3\equiv A_4^3/(2\pi T)$, $a_8\equiv
A_4^8/(2\pi T)$ and $\nu_i = \phi_{i}/(2\pi T)$, \eq{a4ev} becomes
\bea\label{ev}
\pm\nu_1 &=&\pm a_3, \\ \nn
\pm\nu_2 &=& \pm\frac{a_3+a_8\sqrt{3}}{2}, \\ \nn
\pm\nu_3 &=& \pm\frac{a_3-a_8\sqrt{3}}{2}.
\eea
The Polyakov line $L=\exp\left({i A_4/T}\right)$ is a diagonal unitary matrix:
\bea\nn
L &\!\!\!=\!\!\!& 
\left(\begin{array}{ccc}
e^{\pi i\left(a_3 + \frac{a_8}{\sqrt{3}}\right)}&0&0\\
0&e^{-\pi i\left(a_3 - \frac{a_8}{\sqrt{3}}\right)}&0\\
0&0&e^{-\frac{2\pi i}{\sqrt{3}}a_8}\\
\end{array}\right)\\
&\!\!\!=\!\!\!& 
\left(\begin{array}{ccc}
e^{\frac{2\pi i}{3}(\nu_1+\nu_2)}&0&0\\
0&e^{-\frac{2\pi i}{3}(\nu_1+\nu_3)}&0\\
0&0&e^{-\frac{2\pi i}{3}(\nu_2-\nu_3)}\\
\end{array}\right)\!.
\label{Peigenvalues}\eea
Here we expressed it once in terms of $a_3$ and $a_8$ and once through the
eigenvalues of $A_4^{\rm adj}$. 
It is easy to see that the Polyakov line \ur{Peigenvalues} assumes 
values of the group center, $e^{2\pi i k/3}{\bf 1}_{\rm 3},\;k=0,1,2$, 
for integer values of $\nu_i$. This knowledge will be of use 
to understand the potential energy in the next section.

\section{The potential energy}
The potential energy is the contribution to the 1-loop
effective action that corresponds to constant background fields, hence all
spatial variations are set to zero. Since in Yang--Mills theories gauge
invariance requires that derivatives always show up in their covariant form, we obtain the
potential by setting $D_i$ equal to zero. From \eq{S1l} and \eqs{Dpw}{Wpw} we
see that $F_{\mu\nu}$ does not contribute since it is either linear ($F_{4i}$)
or quadratic ($F_{ij}$) in the covariant derivatives. Therefore, 
we find that at zeroth order in $D_i$ 
\bea\label{S1l0}
S_{\rm{ 1-loop}}^{(0)} &\!\!\!=\!\!\!& -\left[\log\,\det \left(-D^2\right)\right]^{(0)} =
\\
\nn
&\!\!\!=\!\!\!&\int d^3 x\skp e^{-sp^2}\Tr \left(e^{-s{\cal A}^2}-e^{-s \omega_k^2}\right) \\ 
\nn
&\!\!\!=\!\!\!&- \int \!d^3x\,P(A_4).
\eea
Here we introduce compact notations:
\bea\nn
\skp &=&\sum_{k=-\infty}^{\infty}\int\frac{d^3 p}{(2
  \pi)^3}\int_0^{\infty}\frac{ds}{s}\,,\\
\nn
\skpm &=& \sum_{k=-\infty}^{\infty}\int\frac{d^3 p}{(2
  \pi)^3}\int_0^{\infty}\frac{ds}{s}\left(1-e^{-s M^2}\right).
\eea
For \eq{S1l0} we need the matrix ${\cal A}^{ab}$. With the 
Ansatz \ur{a4diag} we can construct and
express it in terms of the eigenvalues of $A_4^{\rm adj}$, \eq{a4ev}:
\bea\label{AA}
{\cal A}= \left(
\begin{array}{cccccccc}
i \omega_k & -\phi_1 &0&0&0&0&0&0\\
\phi_1 & i \omega_k  &0&0&0&0&0&0\\
0&0&i \omega_k  &0&0&0&0&0\\
0&0&0&i \omega_k & -\phi_2 &0&0&0\\
0&0&0&\phi_2 &\omega_k &0&0&0\\
0&0&0&0&0&i \omega_k & \phi_3 &0\\
0&0&0&0&0& -\phi_3 &\omega_k &0\\
0&0&0&0&0&0&0&i \omega_k
\end{array}\right).
\eea
We then find that
\bea\nn
&&\Tr\left[e^{-s{\cal A}^2}-e^{-s\omega_k^2}\right]\\
\label{summ}
&=&\!\!\sum_{n=1,2,3}\left[\left(\!e^{-s(\phi_n-\omega_k)^2}\!
+\!e^{-s(\phi_n+\omega_k)^2}\!\right)\! -\! 2 e^{-s {\omega_k}^2}\right]
\eea
where the sum is over the adjoint eigenvalues. In the $SU(2)$ case there
is just one term yielding~\cite{DO2G}
\bea\nn
&&2 \skp\,e^{-sp^2}\left[e^{-s(\phi-\omega_k)^2}-e^{-s\omega_k^2}\right]\\
 &\!\!=\!\!& -\frac{1}{12\pi^2 T}\phi^2(2\pi T -\phi)^2|_{{\rm mod}\;2\pi T}.
\eea
In the $SU(3)$ case one obtains~\cite{GPY} 
\bea\label{V}
P(A_4) = \frac{(2\pi)^2T^3}{3} \sum_{n=1,2,3}\nu_n^2\,(1-\nu_n)^2{ |}_{{\rm mod} \;1},
\eea
where again
\be\label{defnu}
\nu_1=\frac{A_4^3}{2\pi T},\quad \nu_2=\frac{1}{2}\frac{A_4^3+A_4^8\sqrt{3}}{2\pi T}, 
\quad \nu_3=\frac{1}{2}\frac{A_4^3-A_4^8\sqrt{3}}{2\pi T}. 
\ee
We hence see that the potential energy is periodic in the eigenvalues of 
$A_4^{\rm adj}$ with period $2\pi T$. We plot it as a function of $A_4^3$
and $A_4^8$ in Fig.~1. In these axes, the minima and maxima of the potential
energy form a regular triangle lattice, with lattice spacing $2\pi T$. 
The zero-energy minima are at integer values of $\nu_n,\;n\!=\!1,2,3$, where the 
Polyakov line assumes values from the group center. The maxima of the potential
energy correspond to the Polyakov line $L_{\rm max}={\rm diag}
\left(1,e^{\frac{2\pi i}{3}},e^{-\frac{2\pi i}{3}}\right)$ (and those
obtained from this one by multiplying it by the elements of the group center)
whose $\Tr\,L_{\rm max}=0$.    
At high temperatures $L(x)$ oscillates somewhat around
the trivial center-of-group minima of the potential. 
However, as one lowers the temperature the fluctuations of
the Polyakov line around these perturbative minima become stronger and
eventually a transition to a phase with $<\!\Tr\,L\!> = 0$ occurs,
despite the fact that the potential energy is maximal at $\Tr\,L=0$.
The aim of the rest of the paper is to derive the effective 
action for the fluctuations of $L(x)$ in a broad range of its variation 
beyond the trivial minima. 

\begin{figure}[t]
\centerline{
\epsfxsize=0.45\textwidth
\epsfbox{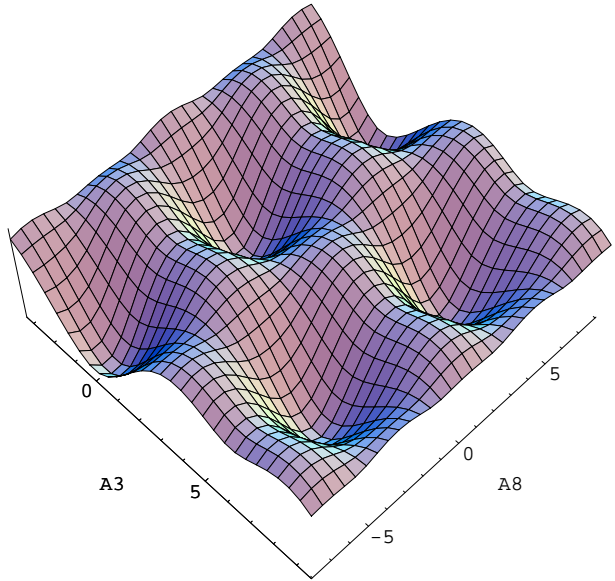}}
\caption{ {\small Potential energy as function of $A_4^3$ and $A_4^8$ forms a 
double-periodic triangle lattice. The $Z(3)$ symmetry is the symmetry 
under lattice translations. }
\label{Pot3D} }
\end{figure}

\section{The electric sector}
As in \cite{DO2G} we are interested in the leading terms for the kinetic
energy in the color-electric sector, meaning terms quadratic in the electric
field. To obtain these, we have to expand \eqs{Dpw}{Wpw} to quadratic order in $D_i$,
because the electric field is identified as \eq{E-B}. Since $D_i$ shows up in
the exponents in \eqs{Dpw}{Wpw} we use the following two `master formulae'
(\cite{DO2G}) for an expansion of exponentials of non-commuting operators:
\bea\label{master}
e^{A+B}\!\! &=&\!\! e^A + \int_0^1 d\alpha e^{\alpha A}B e^{(1-\alpha) A} \\ \nn
&&\,\, + \int_0^1 d\alpha \int_0^{1-\alpha} \!\!\!\!d\beta e^{\alpha
A}B e^{\beta A} B e^{(1-\alpha-\beta)}A + \!\ldots \!\\ \nn
\left[ B, e^A\right]\!\! &=&\!\! \int_0^1 d\gamma e^{\gamma A}
\left[ B,A \right] e^{(1-\gamma)A}.
\eea
Here B contains combinations of $D_i$ and A denotes the rest.
\Eqs{Dpw}{Wpw} as well as \eq{master} are independent of the color
group. We found in \cite{DO2G} that for
any $SU(N)$ the individual contributions from the ghosts and gluons to the
order that we are interested in are
\bea\label{ghqu}
\left[\log\,\det \left(-D^2\right)\right]^{(2)}_{\rm E} \!\!&=&\!\! - \int
d^3 x \skpm 
e^{-sp^2}\left[I_1 + I_2\right],\\ \nonumber
\left[\log\,(\det W)^{-1/2}_{{\rm r,n}}\right]_{{\rm E}}^{(2)}\!\! &=&\!\!
\int d^3
x \skpm e^{-s p^2}\left[\!2 I_1 + 2 I_2 +
\frac{I_3}{2}\!\right],
\eea
so that the 1-loop contribution to the kinetic energy in the electric sector
is
\be\label{SI}
\left[S_{\rm 1-loop}\right]_{\rm E}
=\int\,d^3x\,\skpm\,e^{-s p^2}\left[I_1 + I_2 +\frac{I_3}{2}\right].
\ee
The structures $I_i$ in \eqs{ghqu}{SI} are given by \cite{DO2G}
\bea
\nn
I_1 \!\!&=& \!\!s^3\!\int_0^1 \!d\alpha \! \left\{\!-\frac{1}{2} + \alpha(1\!-\!\alpha)
+\frac{2}{9}sp^2\left[1\!-\!\frac{3}{2}\alpha(1\!-\!\alpha)\right]\!\right\}\\ 
\label{I1}
&&\times
\Tr\,e^{(1-\alpha) s{\cal A}^2}\!\left\{{\cal A}, E_i\right\} \!
e^{\alpha s{\cal A}^2} \!\!\left\{{\cal A}, E_i\right\}\!,\\ 
\label{I2} 
I_2 \!\!&=& \!\!-s^2\!\!\left(\!\! \frac{1}{2}-\frac{2}{9}sp^2\!\!
\right)\!\!\Tr e^{s{\cal A}^2}\left(2 E_i^2 +
i\left\{{\cal A},\left[D_i, E_i\right]\right\}\right),\\
\label{I3}
I_3 \!\!&=& \!\!8s^2\int_0^1\,d\alpha\,\frac{1}{2}\Tr\,e^{(1-\alpha) 
s{\cal A}^2}\,E_i\,e^{\alpha s{\cal A}^2}\,E_i\,,
\eea
where all matrices are in the adjoint representation. 
So far the gauge group has not been specified. For $SU(2)$ the result was
presented in \cite{DO2G}. For $SU(3)$ we
chose $A_4$ to be diagonal in the fundamental representation, \eq{a4diag}, 
and obtain the adjoint matrix ${\cal A}$ in \eq{AA}. The electric field is 
$E_i^{ab}=if^{acb}E_i^{c}$. The second term in the invariant $I_2$ is zero if the gluon
fields obey the equation of motion. We leave it away for the time being but 
shall return to it later. After all integrations and the summation over all Matsubara
frequencies, we find the following structure for the electric kinetic energy:

\bea\label{elresult}
\lefteqn{\left[S_{\rm 1-loop}\right]_{\rm E}=\int\,d^3
x\,\left\{\left[(E_i^1)^2+(E_i^2)^2\right]\lambda_1(\nu)
\right.}
\\
\nonumber
&&\left.
+\left[(E_i^4)^2+(E_i^5)^2\right]\lambda_4(\nu)
+\left[(E_i^6)^2+(E_i^7)^2\right]\lambda_6(\nu)\right.
\\
\nonumber
&&\left.
+(E_i^3)^2\lambda_3(\nu)+
(E_i^8)^2\lambda_8(\nu)+E_i^3 E_i^8 \lambda_{38}(\nu)
\right\}.
\eea
All functions $\lambda_j(\nu)$ are functions of the rescaled
eigenvalues of $A_4^{\rm adj}$, $\nu=\nu_1,\nu_2,\nu_3$, \eq{defnu}. For the
explicit sums over the Matsubara
frequencies, one has to define the range of definition for the $\phi_n$,
respectively the $\nu_n$. This defines the functional form of the
$\lambda_j$. In this section we shall consider the range $0\leq |\phi_i|\leq 2\pi T$, which
corresponds to $0\leq |A_4^3|,\sqrt{3}|A_4^8|\leq 2\pi T$,}. Outside this interval the
functions have to be computed separately. We would like to stress once more 
that we do not expect Z(3)symmetric results here, as we obtained them 
for the potential \ur{V}, which would
be equivalent to invariance under large time-dependent gauge
transformations. We broke this symmetry by making the background
$A_i$ components static. As discussed before, to recover this symmetry
one has to find the action to all orders in time derivatives.

We start by looking at the coefficients in front of the electric field
orthogonal  to $A_4$,  $E_i^{1-2,4-7}$,  namely $\lambda_{1,4,6}$.  We
shall  split  the  results  into  a  contribution  from  the  non-zero
Matsubara frequencies (denoted by a prime) and into the $\omega_{k=0}$
contribution   as  $\lambda_i=\lambda_i^{\prime}+\lambda_{i0}$,  where
$i=1,4,6$. We find that as  long as the individual eigenvalues $\nu_i$
are either  between 0 and  1 or between  -1 and 0 then  the functional
form of the $\lambda_i^{\prime}$ is given by
\bea\nonumber
\!\lambda_1^{\prime}(\nu)\!\! &=&\!\! -\frac{11}{96\pi^2 T}\left[6
  (\ge-\log\,\mu)\! +\! 2\psi\left(\!\frac{\nu_1}{2}\!\right)\! 
  +\! 2\psi\left(\!-\frac{\nu_1}{2}\!\right)\right.\\
\label{lambda1}
&&\left.
      + \psi\left(\!\frac{\nu_2 - \nu_3}{2}\!\right) \!+\! \psi\left(\!-\frac{\nu_2 -
          \nu_3}{2} \!\right)\right],\\\ \nonumber
\!\lambda_4^{\prime}(\nu)\!\! &=&\!\! -\frac{11}{96\pi^2 T}\left[6
  (\ge-\log\,\mu)\! +\! 2\psi\left(\!\frac{\nu_2}{2}\!\right)\! 
  +\! 2\psi\left(\!-\frac{\nu_2}{2}\!\right)\right.\\
\label{lambda4}
&&\left.
      + \psi\left(\!\frac{\nu_1 + \nu_3}{2}\!\right)\! 
      +\! \psi\left(\!-\frac{\nu_1 + \nu_3}{2}\! \right)\right],\\\nonumber
\!\lambda_6^{\prime}(\nu)\!\! &=&\!\! -\frac{11}{96\pi^2 T}\left[6
  (\ge-\log\,\mu)\! +\! 2\psi\left(\!\frac{\nu_3}{2}\!\right)\! 
  +\! 2\psi\left(\!-\frac{\nu_3}{2}\!\right)\right.\\
\label{lambda6}
&&\left.
      + \psi\left(\frac{\nu_1 + \nu_2}{2}\right)\! +\!
      \psi\left(\!-\frac{\nu_1 + \nu_2}{2} \!\right)\right].
\eea
Here $\psi$ is the digamma function
\be
\psi(\nu) = \frac{\partial}{\partial \nu}\,\log\, \Gamma(\nu).
\ee
For the $\lambda_{i0}$ their functional form depends on the region of
definition of each of the three $\nu_i$ separately. It is, however, possible
to express them in terms of the $\phi_i$ before specifying the intervals
for the latter:
\bea\nonumber
\!\lambda_{10} \!\!\!&=&\!\!\! \frac{1}{2\pi^2}\!\left[\!\frac{5\pi}{3 |\phi_1|}\! +\!
  \frac{|\phi_2|\pi\left(10\phi_2^3 + 23 \phi_2^2 \phi_3 + 10 \phi_2 \phi_3^2
      + \phi_3^3\right)}{12\phi_2 (\phi_2-\phi_3)(\phi_2+\phi_3)^3} 
\right.
\\\label{lambda10}
&&\left.-
  \frac{|\phi_3|\pi\left(\phi_2^3 + 10 \phi_2^2 \phi_3 + 23 \phi_2 \phi_3^2 + 10
    \phi_3^3\right)}{12\phi_3(\phi_2-\phi_3)(\phi_2+\phi_3)^3}\right],
\eea
\bea\nonumber
\!\lambda_{40} \!\!\!&=&\!\!\! \frac{1}{2\pi^2}\!\left[\!\frac{5\pi}{3|\phi_2|}\! +\!
  \frac{|\phi_3|\pi\left(-10\phi_3^3 + 23 \phi_3^2 \phi_1 - 10 \phi_3 \phi_1^2
      + \phi_1^3\right)}{12\phi_3 (\phi_1+\phi_3)(\phi_1-\phi_3)^3}
\right.
\\\label{lambda40}
&&\left. +
  \frac{|\phi_1|\pi\left(-\phi_3^3 + 10 \phi_3^2 \phi_1 - 23 \phi_3 \phi_1^2 + 10
    \phi_1^3\right)}{12 \phi_1(\phi_1+\phi_3)(\phi_1-\phi_3)^3}\right],
\eea
\bea\nonumber
\!\lambda_{60} \!\!\!&=&\!\!\! \frac{1}{2\pi^2}\!\left[\!\frac{5\pi}{3|\phi_3|}\! +\!
  \frac{|\phi_1|\pi\left(10\phi_1^3 - 23 \phi_1^2 \phi_2 + 10 \phi_1 \phi_2^2
      - \phi_2^3\right)}{12\phi_1 (\phi_1+\phi_2)(\phi_1-\phi_2)^3}
\right.
\\\label{lambda60}
&&\left. +
  \frac{|\phi_2|\pi\left(\phi_1^3 - 10 \phi_1^2 \phi_2 + 23 \phi_1 \phi_2^2 - 10
    \phi_2^3\right)}{12\phi_2(\phi_1+\phi_2)(\phi_1-\phi_2)^3}\right].
\eea
Note that there is no explicit factor 1/T, since the variables are not
rescaled yet.
One can easily see that these functions can be obtained from each other as
$\lambda_{40} =\lambda_{10}\left(\phi_1\to\phi_2, \phi_2\to -\phi_1\right)$
and $\lambda_{60} =\lambda_{40}\left(\phi_2\leftrightarrow\phi_3\right)$.
To show how the functions (\ref{lambda10}-\ref{lambda60}) depend on the
region of definition of $\phi_i$ we
show some examples. If all three $\nu_i=\phi_i/(2\pi T)$ are defined to be in the interval
from 0 to +1, then we have
\bea\nn
\lambda_{10} \!\!\! &=&\!\!\!\frac{1}{12\pi^2 T}\left[\!\frac{5}{{\nu_1}}\! 
-\! \frac{{\nu_2}^2}{({\nu_2}\!+\!{\nu_3})^3}\! \!+\!\!
  \frac{{\nu_2}}{({\nu_2}\!+\!{\nu_3})^2}\! \!+\!\! \frac{9}{4({\nu_2}\!+\!{\nu_3})}\!\right],\\
\nn
\lambda_{40}\!\!\! &=&\!\!\!\frac{1}{12\pi^2 T}\left[\!\frac{5}{{\nu_2}} \!+\!
  \frac{11}{4({\nu_1}\!+\!{\nu_3})}\!\right],\\
\lambda_{60}\!\!\! &=&\!\!\!\frac{1}{12\pi^2 T}\left[\!\frac{5}{{\nu_3}} \!+\!
  \frac{11}{4({\nu_1}\!+\!{\nu_2})}\!\right].
\eea
If all three $\nu_i$ lie between -1 and 0, then the above functions just
change their global sign. If we put $\nu_1$ and $\nu_3$ in the positive
region from 0 to 1 and $\nu_2$ into the negative one from -1 to 0, then the
functions change to
\bea
\lambda_{10}\!\! &=&\!\!\frac{1}{12\pi^2 T}\left[\frac{5}{{\nu_1}}\!+\!
  \frac{11}{4({\nu_3}-{\nu_2})}\right],\\\nn
\lambda_{40} \!\!&=&\!\!\frac{1}{12\pi^2 T}\left[-\frac{5}{{\nu_2}}\!+\!
  \frac{11}{4({\nu_1}\!+\!{\nu_3})}\right],\\\nn
\lambda_{60}\!\! &=&\!\!\frac{1}{12\pi^2 T}\left[\frac{5}{{\nu_3}} 
- \frac{{\nu_1}^2}{({\nu_1}-{\nu_2})^3} \!+\!
  \frac{{\nu_1}}{({\nu_1}-{\nu_2})^2} \!+\! \frac{9}{4({\nu_1}-{\nu_2})}\right].
\eea
Adding up the contributions from the non-zero Matsubara frequencies,
eqs.(\ref{lambda1}-\ref{lambda6}), and the $\lambda_{i0}$ terms from
eqs.(\ref{lambda10}-\ref{lambda60}), one finds that the functions are not symmetric in
the $\nu_i$ between 0 and 1 (or between -1 and 0), which would have been the
requirement for the Z(3) equivalence. 

The remaining coefficient functions $\lambda_{3,8,38}(\nu)$ in \eq{elresult}, 
which are connected to the electric field parallel to $A_4$, viz. $E_i^3$ and  
$E_i^8$, can be expressed in terms of the function
\bea\label{h}
h(\nu) = \frac{1}{|\nu|} + 2 (\log \mu - \ge) - \psi(\nu) - \psi(-\nu).
\eea
In the positive region $0\le \nu \le 1$ we can use 
that $\psi(\nu)+ 1/\nu = \psi(1+\nu)$ and \eq{h} becomes
\bea
h(\nu)  &=& 2(\log\,\mu -\ge) - \psi(\nu) - \psi(1-\nu)\\ \nn &\equiv& 2\log\,\mu+H(\nu),\\
H(\nu) &=& - \psi(\nu) - \psi(1-\nu)-2\ge.
\eea 
Inside the interval $0\le \nu \le 1$ the function is symmetric with respect
to the replacement $\nu\to 1-\nu$. Outside this interval the function is
continued by periodicity with period 1. We plot the cutoff-independent function
$H(\nu)$ in Fig.~2. We would like to stress that the term $1/\nu$ 
is always the (sole) contribution of the zero Matsubara frequency $\omega_0$. 
  
\begin{figure}[t]
\centerline{
\epsfxsize=0.45\textwidth
\epsfbox{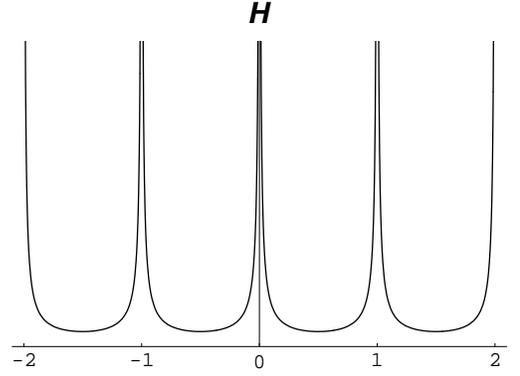}}
\caption{ {\small The periodic function $H(\nu)$ plotted versus $\nu$. }\label{hnu} }
\end{figure}

If we keep all the $\nu_n$ either between -1 and 0 or
between 0 and 1 we find the following functional form for the coefficients in
the parallel sector:
\bea\label{lambda3}
\lambda_3(\nu) &=& \frac{11}{192 \pi^2 T}\left[4 h({\nu_1}) + h({\nu_2}) +
  h({\nu_3})\right],\\\label{lambda8}
\lambda_8(\nu) &=& \frac{11}{64 \pi^2 T}\left[ h({\nu_2}) +
  h({\nu_3})\right],\\\label{lambda38}
\lambda_{38}(\nu) &=& \frac{11}{32\sqrt{3} \pi^2 T}\left[ h({\nu_2}) -
  h({\nu_3})\right].
\eea
They are periodic functions in the eigenvalues of $A_4^{\rm adj}$. 

Similar to the case of $SU(2)$ discussed in \cite{DO2G}, the periodicity 
in the parallel sector comes from the fact that the electric field components
$E_i^{3,8}=\partial_iA_4^{3,8}+f^{(3,8)c(3,8)}A_i^cA_4^{3,8} = \partial_iA_4^{3,8}$
do not dependent on $A_i$ and are by themselves invariant under special 
time-dependent gauge transformation \ur{discreteGT} leading to the shift of
$A_4^{3,8}$ by integers. Since the action has to be invariant under \ur{discreteGT},
it can only happen if the coefficient functions $\lambda_{3,8,38}(\nu)$ are
periodic functions of $\nu_m$ which indeed they are owing to the periodicity
of the functions $h(\nu)$. The rest electric field components necessarily contain
the $A_i$ fields which become fast time-dependent under the transformation \ur{discreteGT},
see eqs.(\ref{Ai1}-\ref{Ai7}). The full electric field is $E_i^a=D_i^{ab}A_4^b-\dot A_i^a$.
Therefore, the transformation \ur{discreteGT} changes the components $E_i^{1,2,4,5,6,7}$,
and the invariance under \ur{discreteGT} does not request that the corresponding
coefficient functions $\lambda_{1,4,6}(\nu)$ are periodic. The $Z(3)$ symmetry in the
`transverse' electric sector has to take place only when one collects all powers
of $\dot A_i$.

All functions $\lambda_1,\ldots,\lambda_8$ (but not the cross-term
$\lambda_{38}$) contain the same UV divergent contribution
\be\label{UV}
\frac{11}{16\pi^2 T}\log\,\mu.
\ee
The parameter $\mu$ is the UV-cutoff in divergent series:
\be
\sum_{k=1}^{\infty}\frac{1}{k}\to\sum_{k=1}^{\mu}\frac{1}{k}\equiv \log\,\mu \,,
\ee
and is related to the Pauli--Villars mass as
\bea
\mu=\frac{M}{4\pi T}e^{\ge}.
\eea
This subtraction scale for the running coupling constant has been known
previously \cite{coupling} and was also obtained in \cite{DO2G}.

The divergence \ur{UV} is necessary to cancel the UV divergence from the running coupling
constant in the tree level action:
\bea\label{run}
-\frac{F_{\mu\nu}^2}{4 g^2(M)} &=&
-\frac{F_{\mu\nu}^2}{8\pi^2}\log\frac{M}{\Lambda}\frac{11}{12}
  N_c \\ \nn
&=& - (E_i^2+B_i^2)\frac{11}{16\pi^2}\log\frac{M}{\Lambda}.
\eea
If we add the tree-level  and the 1-loop action then the
result should be UV finite. This is obtained by choosing the
scale $M$ in \eq{run} to be equal to the Pauli--Villars mass, 
which corresponds to the evaluation of the running coupling constant 
at the scale $4\pi T/\exp(\ge)$. In the effective action we then have
to replace the Pauli-Villars cutoff $M$ by $\Lambda$ in all our functions
$\lambda_i$ from \eq{elresult}.

\section{Gauge invariant structures}
Written in the form above the 1-loop action \eq{elresult}
does not look covariant. This is, however, a consequence of the fact that we
chose $A_4$ to be diagonal. It is easy to show that \eq{elresult} can be
rewritten as
\bea\label{Selinv}
\left[S_{\rm 1-loop}\right]_{\rm E}=\int\,d^3
x\,\sum_{i=1}^5 {\Phi}_i f_i(\nu_1, \nu_2, \nu_3)
\eea
with the following five invariant structures:
\bea\label{strucs}
{\Phi}_1 &=& \Tr [E_i A_4]\Tr [E_i A_4] \\\nonumber
{\Phi}_2 &=& \Tr [E_i E_i] \\\nonumber
{\Phi}_3  &=& \Tr [E_i A_4 E_i A_4] \\\nonumber
{\Phi}_4  &=& \Tr [E_i {A_4}^2 E_i {A_4}^2]\\\nonumber
{\Phi}_5  &=& \Tr [E_i {A_4}^3] \Tr [E_i {A_4}^3] 
\eea
All the traces are over matrices in the
adjoint representation. For our particular choice \ur{a4diag} for $A_4$ these
structures are composed as follows:
\bea\label{strucex}
{\Phi}_1 \!\!&=&\!\! 4\pi^2 T^2 \!\left[\!\sqrt{3} E_i^8 ({\nu_2}-{\nu_3})
 \! +\! E_i^3 (2 {\nu_1}+{\nu_2}+{\nu_3})\right]^2,\\\nonumber
{\Phi}_2\!\! &=&\!\! 3\!
\left[(E_i^a E_i^a)\right],\\\nonumber
{\Phi}_3 \!\! &=&\!\! 2 \pi^2 T^2\!
\left[ -2 {\nu_2}
    {\nu_3}({E_i^1}^2\!+\!{E_i^2}^2)\!+\!2 {\nu_1}
    {\nu_3}({E_i^4}^2\!+\!{E_i^5}^2)
\right.
\\\nonumber
&&\left.
+2 {\nu_1} {\nu_2}({E_i^6}^2\!+\!{E_i^7}^2)\!+\! {E_i^3}^2
  (4 {\nu_1}^2 \!+\! {\nu_2}^2 \!+\! {\nu_3}^2) 
\right.
\\\nonumber
&&\left.
+ 3{E_i^8}^2({\nu_2}^2 \!+\! {\nu_3}^2) \!+\!
  2\sqrt{3} E_i^3 E_i^8 ({\nu_2}-{\nu_3})\right],\\\nonumber
{\Phi}_4 \!\! &=&\!\! 8 \pi^4 T^4\!
\left[ 2 {\nu_2}^2
    {\nu_3}^2({E_i^1}^2\!+\!{E_i^2}^2)\!+\!2 {\nu_1}^2
    {\nu_3}^2({E_i^4}^2\!+\!{E_i^5}^2)
\right.
\\\nonumber
&&\left.
+2 {\nu_1}^2 {\nu_2}^2\left({E_i^6}^2\!+\!{E_i^7}^2\right)\!+\! {E_i^3}^2
  (4 {\nu_1}^4 \!+\! {\nu_2}^4 \!+\! {\nu_3}^4)\right.
\\\nonumber
&&\left.
 + 3{E_i^8}^2({\nu_2}^4 \!+\!{\nu_3}^4) \!+\!
  2\sqrt{3} E_i^3 E_i^8 ({\nu_2}^4-{\nu_3}^4)\right],\\\nonumber
{\Phi}_5 \!\! &=&\!\! 64 \pi^6 T^6
\left[\sqrt{3} E_i^8 ({\nu_2}^3-{\nu_3}^3)
  \!+\! E_i^3 (2 {\nu_1}^3 \!+\! {\nu_2}^3 \!+\! {\nu_3}^3)\right]^2.
\eea
Since the coefficients $f_i$ in front of these structures in \eq{Selinv}, which reproduce
\eq{elresult}, are rather lengthy we do not write them down explicitly here. But it is easy to
reconstruct them with \eq{strucex}. 
We would like to stress, however, that the UV logarithm is now only contained
in the coefficient $f_2$ in front of ${\Phi}_2$. This is expected, since only that
structure shows up in the classical action \eq{run}.

\section{The magnetic sector}
To get the magnetic fields in the effective action we have to
expand \eqs{Dpw}{Wpw} to the quartic order in $D_i$.
The basic idea of the calculation is, again, to use master equations
\ur{master} to drag covariant derivatives $D_i$
to the right. What enters in the expansion is 
\bea
D_i e^{s{\cal A}^2} &\!\!\!=\!\!& e^{s{\cal A}^2}\! D_i\! -\! is\!\!\int_0^1\!
d\delta e^{\delta s{\cal A}^2}\{{\cal A},E_i\}e^{(1-\delta) s{\cal A}^2},\\ \nn
D_i D_j\,e^{s{\cal A}^2} &\!\!\!=\!\!& e^{s{\cal A}^2}\! D_i
D_j\! -\! is\!\!\int_0^1\!\! d\delta e^{\delta s{\cal A}^2}[D_i
D_j, {\cal A}^2]e^{(1-\delta) s{\cal A}^2},
\eea
where
\begin{equation}
[D_i D_j, {\cal A}^2] = -i D_i\{{\cal A},E_j\} -
i\{{\cal A},E_i\}D_j.
\nonumber\end{equation}

In this way one ultimately obtains gauge-invariant combinations of the
electric  field  in the  fourth  power,  mixed  terms containing  both
electric and  magnetic fields, derivatives of the  electric field and,
finally, magnetic  field squared. In this paper  we restrict ourselves
to the latter terms quadratic  in the magnetic field $B_i$. This means
that we treat  $D_i$ and $A_4$ as commuting  operators and neglect the
commutators $[D_i, A_4]$.  Moreover, unlike for the electric field, we
would  like to  restrict ourselves  to  a magnetic  field parallel  to
$A_4$,  i.e.   $B_i=B_{i\parallel}$. In  that  case  we  see from  the
Bianchi   identity  $[F_{ij},A_4]=i\,\left([D_i,E_j]-[D_j,E_i]\right)$
that we can neglect derivatives  of the electric field as well.  Hence
we can also treat $D_i$ and $F_{ij}$ as commuting operators.

As shown in  \cite{DO2G} there is only one  structure that contributes
to quadratic order in the magnetic field:
\bea\label{Smag}
\lefteqn{\left[S_{\rm{1-loop}}^{(2)}\right]_{{\rm M}} 
= \left[\log\, \left(\det{W}\right)^{-1/2}_{{\rm r,n}} +
  \log\,\det \left({-D^2}\right)_{{\rm r,n}}\right]^{(2)}_{{\rm M}}  }\\ 
\nn 
&& \!\!= \!\!
\frac{11}{48\,\pi^{3/2}} \!\int  \!d^3 x  \!\!\!\sum_{k=-\infty}^{\infty} \!\!
\int_0^{\infty} \!\!\frac{ds}{\sqrt{s}} \!\Tr\left( \!1-e^{-s M^2} \!\right)
\left( \!e^{s{\cal A}^2} \! B_k B_k \! \right)\,.
\eea
\Eq{Smag} is so far independent of the gauge group. For $SU(3)$ we have the
matrix ${\cal A}$ given by \ur{AA}, and the magnetic field is in the adjoint
representation, $B_i^{ab}=if^{acb}B_i^c$ where $B_i^c$ is as in \eq{E-B}. 
After the integrations over $s$ and $p$ and the summation over the Matsubara
frequencies $\omega_k$ we find the following result:
\bea\label{magresult}
&&\left[S_{\rm 1-loop}\right]_{\rm M}=\int d^3x
\left\{(B_i^3)^2\kappa_3(\nu)  \right. \\ \nonumber && \left. \quad +
(B_i^8)^2\kappa_8(\nu)+B_i^3 B_i^8 \kappa_{38}(\nu)
\right\}.
\eea

Again  the  coefficients  $\kappa_i$  are functions  of  the  rescaled
eigenvalues  of $A_4^{\rm  adj}$,  $\nu=\nu_1,\nu_2,\nu_3$, and  their
functional form  will depend on the  region of definition  of them. As
before we chose $0\leq |\phi_i|\leq  2\pi T$. The fact that the result
\ur{magresult}  looks  non-invariant is  again  a  consequence of  the
diagonal form  of $A_4$. However,  \eq{magresult} can be  rewritten in
terms  of  the  invariants  \ur{strucs} with  an  obvious  replacement
$E_i^{3,8}\to B_i^{3,8}$ and  $E_{i}^{1,4,6}\to 0$.  All the functions
$\kappa_j$  are linear combinations  of the  function $h$,  defined in
\eq{h}, with arguments $\nu_{1,2,3}$:
\bea\nn
\kappa_3(\nu) &\!\!=\!\!& \frac{11}{192 \pi^2 T}\left[4 h({\nu_1}) + h({\nu_2}) +
  h({\nu_3})\right]= \lambda_3(\nu),\\\nonumber
\kappa_8(\nu) &\!\!=\!\!& \frac{33}{192 \pi^2 T}\left[h({\nu_2}) + h({\nu_3})\right]
= \lambda_8(\nu),\\
\label{kappa}
\kappa_{38}(\nu) &\!\!=\!\!& \frac{11}{32\sqrt{3}\, \pi^2 T}\left[h({\nu_2}) 
- h({\nu_3})\right]= \lambda_{38}(\nu).
\eea
All of these functions are symmetric in the rescaled eigenvalues
$\nu$ of $A_4^{\rm adj}$, which corresponds to the equivalence of 
the Z(3)-symmetric points. The reason for this is similar to 
the one we discussed for the parallel electric sector. Since the magnetic field
\be
B^a_i=\frac{1}{2}\epsilon_{ijk}\left(\partial_jA_k^a-\partial_kA_j^a
+\epsilon^{abc}A_j^bA_k^c\right)
\ee
does not have any explicit time dependence, a fast time-dependent gauge
transformation does not affect ${B_i}^2$, neither for static nor for time dependent
$A_i$ fields. The invariance of the action with respect to
the special time-dependent gauge transformation \ur{discreteGT} 
requires then that \eq{magresult} is $Z(3)$-symmetric, which amounts to the
periodicity of all functions $\kappa(\nu)$. Indeed, they are all periodic. 

The functions $\kappa_3,\kappa_8$ have the same UV divergent
contribution 
\be
\frac{11}{16\pi^2 T}\log\,\mu ,
\ee
while the cross-term $\kappa_{38}$ is UV finite. This divergence removes the
one from the tree level action, \eq{run}. Adding \eq{run} and \eq{magresult}
corresponds to the evaluation of the running coupling constant 
at the scale $4\pi T/\exp(\ge)$. In the effective action we then have
to replace the Pauli-Villars cutoff $M$ by $\Lambda$ in all our functions
$\kappa_i$ from \eq{magresult}.

\section{Invariance under permutations of color axes}
An important check of the results \urs{elresult}{magresult} is
their invariance under color rotations corresponding to the permutation 
of color axes.  We can for example rotate from the color-1 axes to 
the color-4 axes, or from the color-4 axes to the color-6 axes. 
This corresponds to finding the matrices $S$ and $U$ such that:
\bea
S^{\dagger}\lambda^1 S &=& \lambda^4,\\
U^{\dagger}\lambda^4 U &=& \lambda^6.
\eea
They are given by 
\bea\label{SU}
S= \left(\begin{array}{ccc}
1&0&0\\0&0&1\\0&-1&0
\end{array}\right)\quad,\quad
U=\left(\begin{array}{ccc}
0&1&0\\-1&0&0\\0&0&1
\end{array}\right).
\eea
We shall show explicitly how this works for the first of these two rotations,
namely the one with $S$ around $\lambda^1$. We chose $A_4$ to be diagonal in
the fundamental representation, \eq{a4diag},
\be
A_4 = A_4^3 t^3 + A_4^8 t^8.
\ee
Under the rotation $S^{\dagger}A_4 S$ we find the transformations
\bea\label{rotateA4}
A_4^3 \to \frac{A_4^3 + \sqrt{3}A_4^8}{2}\equiv\tilde{A_4^3}\,,\, A_4^8 \to
\frac{3 A_4^3-
  \sqrt{3}A_4^8}{2\sqrt{3}}\equiv\tilde{A_4^8}.
\eea
The electric and magnetic fields transform with ${\cal
  S}^{ab}=\half\Tr\left(S^{\dagger}\lambda^a S \lambda^b\right)$ to
\bea\label{rotateEB}
\lefteqn{{\cal S}^{ab}E_i^b \equiv
\tilde{E_i^a}}\\\nn
&&\!\!\!\!=\left(\!\!E_i^4, E_i^5, \!\frac{E_i^3\! + \!\sqrt{3}E_i^8}{2},
 \! -E_i^1,\!-E_i^2,\!-E_i^6,E_i^7,\!\frac{\sqrt{3}E_i^3\! -\!E_i^8}{2}
 \!\!\right),
\eea
\bea
\lefteqn{{\cal S}^{ab}B_i^b \equiv
\tilde{B_i^a}}\\\nn
&&\!\!\!\!=\left(\!\!0, 0, \frac{B_i^3 + \sqrt{3}B_i^8}{2},
 0,0,0,0,\frac{\sqrt{3}B_i^3 -B_i^8}{2}
 \!\!\right).
\eea
We hence see that the structures orthogonal and parallel to $A_4$ transform
separately. 
We shall start by checking the results in the electric sector. For the
orthogonal part we have to show that
\bea\label{rotatelambda}
\lambda_1(A_4^3, A_4^8) &=& \lambda_4(\tilde{A_4^3}, \tilde{A_4^8}),\\\nn
\lambda_4(A_4^3, A_4^8) &=& \lambda_1(\tilde{A_4^3}, \tilde{A_4^8}),\\\nn
\lambda_6(A_4^3, A_4^8) &=& \lambda_6(\tilde{A_4^3}, \tilde{A_4^8}),
\eea
and for the parallel part the requirement is
\bea\nn
&&{E_i^3}^2\lambda_3(A_4^3, A_4^8) +{E_i^8}^2\lambda_8(A_4^3, A_4^8) + E_i^3
E_i^8 \lambda_{38}(A_4^3, A_4^8)\\\label{rotatePar}
&& =\tilde{E_i^3}^2 \lambda_3(\tilde{A_4^3},
\tilde{A_4^8})\! +\! \tilde{E_i^8}^2 \lambda_8(\tilde{A_4^3},
\tilde{A_4^8}) \\ \nn
&&\,\,\,\,\,\, +  \tilde{E_i^3}\tilde{E_i^8}\lambda_{38}(\tilde{A_4^3},\tilde{A_4^8}) .
\eea
There is one additional subtlety, however. Here we expressed the functions
through $A_4^3,A_4^8$, instead of through the eigenvalues $\nu_i$. We
mentioned before that the functional form of the $\lambda_i$ depends on the
interval of definition for the individual $\nu_i$. So if we compare functions
of $A_4^3,A_4^8$ to functions of ${\tilde A_4^3},\tilde{A_4^8}$, we have to
classify these functions according to their support in $\nu_i$, i.e. the
$\nu_i$'s that correspond to $A_4^3,A_4^8$ can lie in different intervals
than the $\nu_i$'s corresponding to ${\tilde A_4^3},\tilde{A_4^8}$. For a
comparison we then have to choose different functional forms of the $\lambda_i$.

It turns out, however, that as long as $A_4^3/(2\pi T)$ and $\sqrt{3}A_4^8/(2\pi T)$
are between -1 and 1 (then also the $\nu_i$ are in that region) we can check
the contribution of the zero Matsubara mode and of the non-zero modes
separately. 

We start by checking the contribution of the zero Matsubara mode in the
orthogonal electric sector, eqs.(\ref{lambda10}-\ref{lambda60}). 
It is independent of the region of definition of $A_4^3/(2\pi T)$ 
and $A_4^8/(2\pi T)$, since it is not rescaled. First we express it in terms of
$A_4^3$ and $A_4^8$, then we make the rotation to $\tilde{A_4^3},
\tilde{A_4^8}$ according to \eq{rotateA4}. And we find indeed that
$\lambda_{10}(A_4^3, A_4^8) = \lambda_{40}(\tilde{A_4^3}, \tilde{A_4^8})$,
$\lambda_{40}(A_4^3, A_4^8) = \lambda_{10}(\tilde{A_4^3}, \tilde{A_4^8})$ and
$\lambda_{60}(A_4^3, A_4^8) = \lambda_{60}(\tilde{A_4^3},
\tilde{A_4^8})$. The $\lambda_i^{\prime}$ in eqs.(\ref{lambda1}-\ref{lambda6}) 
have the same functional form for individual $\nu_i$ in the positive or 
negative region. Therefore, we first express them in terms of
$A_4^3/(2\pi T)$ and $A_4^8/(2\pi T)$ and then make the rotation
\eq{rotateA4}. As expected we find that $\lambda_{1}^{\prime}(A_4^3, A_4^8) 
= \lambda_{4}^{\prime}(\tilde{A_4^3}, \tilde{A_4^8})$,
$\lambda_{4}^{\prime}(A_4^3, A_4^8) = \lambda_{1}^{\prime}(\tilde{A_4^3}, \tilde{A_4^8})$ 
and $\lambda_{6}^{\prime}(A_4^3, A_4^8) = \lambda_{6}^{\prime}(\tilde{A_4^3},
\tilde{A_4^8})$. Therefore, the electric sector orthogonal to $A_4$ is indeed
invariant under color rotations. 

The functional form of eqs.(\ref{lambda3}-\ref{lambda38}) for the 
electric field parallel to $A_4$ depends on the region of definition
of each of the three $\nu_n$. It is again the contribution of the zero
Matsubara frequency which is responsible for that. It comes in as the $1/|\nu|$
term in the definition of the function $h(\nu)$ in \eq{h}. If $0\le \nu \le 1$
then one has $+1/\nu$, while it becomes $-1/\nu$ if $-1\le \nu \le 0$. As discussed
above, however, the color rotation \eq{rotateA4} can rotate individual
$\nu_n$ from a positive region of definition into the negative one, and vice
versa. This problem can be circumvented, however, by replacing the argument
of the function $h(\nu_n)$ in the coefficients
$\lambda_{3,8,38}(\nu)$ of eqs.(\ref{lambda3}-\ref{lambda38}) by the fractional part
of $\nu_n$. Then it is easy to check that the invariance requirement \eq{rotatePar} is
fulfilled for each value of $A_4^3$ and $\sqrt{3}A_4^8$ between $-2\pi T$ and $2\pi
T$, where the functions $h(\nu_n)$ are defined.

What remains is to check the results for the magnetic sector of the theory,
\eqs{magresult}{kappa}. Since the invariance requirements
\eqs{rotatelambda}{rotatePar} remain the same in the electric sector if we
replace $\lambda_i\to \kappa_i$ and $E_i \to B_i$ we have already shown the
invariance of the magnetic sector parallel to $A_4$, since according to \eq{kappa}
$\kappa_{3,8,38}=\lambda_{3,8,38}$. 

Similarly one checks the invariance under other color rotations, for example
the one with $U$ in \ur{SU}.

\section{The ``equation of motion'' term}

We finally return to the second term in the invariant $I_2$,
\eq{I2}. In the derivation of our results in the previous sections we ignored
this term since it vanishes if the gluon fields obey the equation of
motion. Its contribution to the effective action is
\bea\label{eof}
\lefteqn{S_{{\rm EoM}} =\int d^3 x \sum_{k=-\infty}^{\infty}\int\frac{d^3 p}{(2
\pi)^3} \int_0^{\infty}\frac{ds}{s} e^{-s p^2}}\\ \nonumber
&&\times \left[-s^2\left( \frac{1}{2}-\frac{2}{9}sp^2 \right)
\Tr e^{s{\cal A}^2}\left(i\left\{{\cal A},\left[D_i, E_i\right]\right\}\right)\right]\;,
\eea
which after integrations and summation yields
\be\label{eom}
S_{{\rm EoM}} = \int d^3 x \frac{1}{8\pi}\left[\frac{(D_i E_i)^a\,A_4^a}{\pi T} - \frac{(D_i
E_i)^b A_4^b}{\sqrt{A_4^c A_4^c}}\right] .
\ee
Here the first term comes exclusively from the nonzero Matsubara
frequencies, while the second term is the contribution of the zero Matsubara
frequency alone.  The result is similar to the one that we obtained for the
case of $SU(2)$. The only difference is the prefactor: while it is $1/8$ here,
it was $1/12$ in \cite{DO2G}. Equation (\ref{eom}) is zero if the classical equation of
motion is satisfied. If the background field does not satisfy the equation of
motion one can integrate eq. (\ref{eom}) by parts which yields:
\bea
\lefteqn{S_{{\rm EoM}} = \frac{1}{8\pi^2} \int\!d^3 x\left\{E_i^a
     E_i^a\left(\frac{\pi}{|A_4|} - \frac{1}{T}\right) \right.}\\ \nonumber 
&&\left.+
     \frac{\pi}{|A_4|^3}\left(E_i^a A_4^a\right)\left(E_i^b A_4^b\right) -
     \partial_i\left(E_i^a A_4^a\left(\frac{\pi}{|A_4|} -
         \frac{1}{T}\right)\right)\right\}.
\eea
The last term here is a full derivative, and the remaining two have to be
added to our results found before in the electric sector. 

\section{Comparison with previous work}
In a related work by Chapman \cite{chapman} an effective action for the 
static modes in a pure Yang--Mills theory was obtained by means of a 
covariant derivative expansion. While we kept all orders of $A_4$ 
in the present work the author of \cite{chapman} expands to quadratic order in
the $A_4$ fields. For a comparison we hence have to expand our functions to
quadratic order in $A_4$. The result for the electric sector is
\bea\nn
&&\!\!\!\!\frac{11\zeta (3)}{768\pi^2 T}\left\{\!2\frac{A_4^3 A_4^8}{(2\pi
    T)^2}\left[\!-\!{E_i^4}^2\!-\!{E_i^5}^2 \!+\!{E_i^6}^2 \!+\! {E_i^7}^2 
\!+\! 8\sqrt{3}E_i^3 E_i^8 \right] 
\right. \\ 
\nn
&& \qquad\left. 
\!+\!\frac{(A_4^3)^2}{(2\pi T)^2}\left[8\left(\!{E_i^1}^2\!+\!{E_i^2}^2\!\right) 
\!+\! 36
    {E_i^3}^2  \!+\! 12 {E_i^8}^2
\right.\right. \\ \nn
&& \qquad\qquad\left.\left.
\!+\! 11 \left(\!{E_i^4}^2\!+\!{E_i^5}^2 \!+\!{E_i^6}^2 \!+\! {E_i^7}^2\!\right) \right]\right.\\
\nonumber
&&\qquad\left.\!+\!\frac{(A_4^8)^2}{(2\pi T)^2}\left[4\left(\!{E_i^1}^2
\!+\!{E_i^2}^2\!+\!{E_i^3}^2\!\right)\!+\! 12 {E_i^8}^2 \right.\right. \\ 
\label{chapmanEl}
&& \qquad\qquad\left. \left.
\!+\! 3 \left(\!{E_i^4}^2\!+\!{E_i^5}^2 
\!+\!{E_i^6}^2 \!+\! {E_i^7}^2\!\right)\right]
\!\right\}\!,
\eea
while in the magnetic sector we find 
\bea\label{chapmanMag}
&&\frac{11\zeta (3)}{384\pi^2 T}\left\{8\sqrt{3}\frac{A_4^3 A_4^8}{(2\pi T)^2}
\,B_i^3 B_i^8 \right.\\ 
\nonumber &&\left.
+\frac{(A_4^3)^2}{(2\pi T)^2}\left[ 18
    {B_i^3}^2 + 6 {B_i^8}^2 
\right]+\frac{(A_4^8)^2}{(2\pi
    T)^2}\left[2{B_i^3}^2+ 6 {B_i^8}^2\right]\right\}.
\eea
This agrees with the results found in \cite{chapman} if the gauge group is
chosen to be $SU(3)$ and the magnetic field is restricted to be parallel to $A_4$.

While this paper was in preparation, ref. \cite{arriola} appeared on the web, where a similar expansion in covariant derivatives of the effective action at high temperatures was performed, see also \cite{arriola2}. Many of our results coincide with theirs: in the electric sector the terms in the effective action where the electric field is parallel to the $A_4$ field, i.e. our functions $\lambda_{3,8,38}$, and our results in the (parallel) magnetic sector, $\kappa_{3,8,38}$. While we consider a parallel magnetic field, the authors of \cite{arriola} have a general result. 
The corresponding functions in ref. \cite{arriola} are periodic in $A^{3,8}_4$, but get an infinite contribution from the zero Matsubara frequencies.

\section{Effective action for the eigenvalues of the Polyakov line}
\subsection{SU(3) case}

In the previous sections, we have computed the 1-loop effective action for the
Polyakov line $L(x)=\exp(iA_4(x)/T)$ as a slowly varying element of $SU(3)$. 
Since $L(x)$ rotates under spatial gauge transformations \ur{GT} one has to
introduce a nonzero $A_i$ background field in order to maintain gauge invariance
of the effective action under spatial gauge transformations, and expand it in 
covariant derivatives $D_i$. If, however, only time-independent $A_i$'s are 
introduced, as it is required by static gauge invariance, the $Z(3)$ symmetry
of the effective action is generally lost, as anticipated in section II and
checked in section V. In order to restore the $Z(3)$ symmetry of the effective action,
one needs to sum up all time derivatives of $A_i$, since the $Z(3)$ symmetry
is a consequence of the invariance under fast time-dependent gauge transformations
\ur{discreteGT}. This seems to be a formidable task which lies beyond the scope 
of this study; therefore we have to restrict ourselves to purely static $A_4,A_i$ 
fields at the cost that the $Z(3)$ symmetry of our effective action is not manifest.           

One can, however, be interested in another problem, namely in the effective quantum
action for the {\em eigenvalues} of the Polyakov line. Contrary to $A_4$ and to the
Polyakov line as a unitary matrix, its eigenvalues {\em are invariant} under spatial gauge
transformations. Therefore, the effective action for the eigenvalues can be expanded
in ordinary `short' rather than covariant derivatives, and the background $A_i$ field
may then be set to zero. In this setting, the $A_i$ fields, as well as
the rapidly changing components of the $A_4$ fields, are understood and treated 
as quantum fluctuations over which one integrates. 

Fortunately, one does not need to solve this problem anew: in fact the 
result can be obtained from the more general case considered above. 
Assuming the gauge where $A_4$ is static and diagonal (such that the eigenvalues
of the Polyakov line are given by \eq{Peigenvalues} expressed through $\nu_{1,2,3}$) 
all one practically needs to do is to set $A_i$ to zero, which means putting
to zero all components of the magnetic field $B_i$ and the non-diagonal components 
of the electric field $E_i^{1,2,4,5,6,7}$. What is left are the $E_i^{3,8}$ components
of the electric field whose definition is 
$E_i^{3,8}=\partial_iA_4^{3,8}+f^{(3,8)c(3,8)}A_i^cA_4^{3,8} = \partial_iA_4^{3,8}$,
i.e. they are independent of $A_i$. One can easily check that these components 
correspond to combinations of the invariants \urs{strucs}{strucex} in which $A_i$
is canceled, even if one does not assume $A_4$ to be diagonal. We remark that
our coefficient functions of $A_4$ in front of $E_i^{3,8}$ coincide with those
computed recently in ref.~\cite{arriola}. 

In terms of the quantities $\nu_{1,2,3}(x)$ the diagonal components of the electric field
are, according to \eq{defnu},
\be
E_i^3=\partial_i\nu_1\,2\pi T,\qquad 
E_i^8=\frac{\partial_i(\nu_2-\nu_3)}{\sqrt{3}}\,2\pi T. 
\label{Ediagonal}\ee
The last line in \eq{elresult} gives the needed 
two-derivative terms in the gauge-invariant effective action for the eigenvalues 
of the Polyakov line. Assembling terms proportional to $h(\nu_1),h(\nu_2),h(\nu_3)$
and using $\nu_1=\nu_2+\nu_3$ we get a remarkably symmetric expression: 
\bea\label{eigenaction}
S^{\rm kin}&\!\!\!=\!\!\!&\int d^3x\,T\,\frac{11}{12}\,\left\{
-2\log\left(\frac{4\pi T}{\Lambda\,e^{\gamma_E}}\right)\right.\\ 
\nn 
&&\left.\!\!\!\!\!\!\times
\left[(\partial_i\nu_1)^2\!+\!(\partial_i\nu_2)^2\!+\!(\partial_i\nu_3)^2\right]\right.\\ 
\nn 
&&\left.\!\!\!\!\!\!+\left[(\partial_i\nu_1)^2H(\nu_1)\!
+\!(\partial_i\nu_2)^2H(\nu_2)\!+\!(\partial_i\nu_3)^2H(\nu_3)\right]\right\},
\eea
where
\bea\label{H1}
H(\nu)=\left[ -\psi(\nu)-\psi(1-\nu)-2\gamma_E\right]_{\rm mod\; 1},
\eea
and the eigenvalues of the Polyakov loop in terms of $\nu_{1,2,3}(x)$ are given by \eq{Peigenvalues}.  
Since the $Z(3)$ symmetry consists in shifting $\nu_{1,2,3}$ (i.e. the eigenvalues of $A_4$ 
in the adjoint representation) by integers, and the function $H(\nu)$ is periodic with 
period 1, this action is explicitly $Z(3)$ symmetric. It should be reminded that $\nu_{1,2,3}$
are not independent but satisfy $\nu_1=\nu_2+\nu_3$, see \eq{defnu}. The renormalization
has been performed in the Pauli--Villars scheme. If another, e.g. the Modified Minimal Subtraction ($\overline{\rm MS}$) scheme
is used, one has to replace $\Lambda$ by $e^{\frac{1}{22}}\Lambda_{\overline{\rm MS}}$ \cite{Has}.   

The function $H(\nu)$ (which appears also in the $SU(2)$ case \cite{BGK-AP,DO2G}) is plotted 
in \fig{hnu}. It goes as $\frac{1}{\nu}$ or as $\frac{1}{1-\nu}$ as $\nu$ 
approaches 0 or 1, respectively. The singularity is due to the contribution 
of the zero Matsubara frequency, i.e. of the static quantum fluctuations. 
One may wish to disregard this contribution but then the periodicity of $H(\nu)$ 
is lost, together with the $Z(3)$ symmetry of the action.

\Eq{eigenaction} together with \eq{V} for the potential energy is the $SU(3)$ effective
action for the eigenvalues of the Polyakov line \ur{Peigenvalues}, up to two
spatial derivatives in the 1-loop order. It is interesting that both the `potential'
and `kinetic' energy parts are sums over eigenvalues of $A_4$ in the adjoint representation. 

\subsection{General SU(N) case}

As before, it is possible and convenient to choose the gauge such that $A_4$ is
static and diagonal in the fundamental representation:
\bea\label{A4N}  
A_4(x)&=&A_4^at^a=2\pi T\;{\rm diag}(a_1,a_2,...,a_N),\\
\nn
&&a_1+a_2+...+a_N=0,
\eea
where the common factor $2\pi T$ is taken for future convenience.
The Polyakov line is then a diagonal $SU(N)$ unitary matrix,
\be\label{PolN}
L(x)={\rm diag}\left(e^{2\pi i a_1},e^{2\pi i a_2},...,e^{2\pi i a_N}\right).
\ee
The effective action must be invariant under time-dependent gauge transformations, 
generalizing \eq{discreteGT}:
\bea\label{SS}
A_4&\!\!\to \!\!& SA_4S^\dagger+iS\partial_t S^\dagger,\\
\label{discreteGT_N}
S&\!\!=\!\!&{\rm diag}\!\left(\!e^{i[2\pi tT \mu_1+\alpha_1(x)]},\ldots,
e^{i[2\pi tT \mu_N+\alpha_N(x)]}\!\right)\!,\\
\label{condition1}
&&\sum_{m=1}^N \mu_m=0,\qquad \sum_{m=1}^N \alpha_m(x)=0. 
\eea
This gauge transformation amounts to shifting $a_m\to a_m+\mu_m$.
The time frequencies $\mu_m$ are quantized because this gauge transformation
written down for the adjoint representation must be periodic in time,
meaning that $O^{ac}(t\!=\!0)O^{bc}(t\!=\!1/T)=\delta^{ab}$ where
$O^{ab}=2\Tr(St^aS^\dagger t^b)$, $\Tr\,t^at^b = \half\delta^{ab}.$
This periodicity in time leads to the requirement that all differences
\be\label{condition2}
\mu_m-\mu_n={\rm integers}.
\ee
Given the condition \ur{condition1}, the general solution to \eq{condition2} is
\be\label{condition3}
\mu_m=\frac{k_0}{N}+k_m,\qquad \sum_{m=0}^Nk_m=0,
\ee
where $k_0,k_1,...,k_N$ are all integer numbers. Under the gauge
transformation \ur{discreteGT_N} the Polyakov line is multiplied by
a diagonal matrix 
\be
S^\dagger(0)S(1/T)={\rm diag}\left(e^{2\pi i \mu_1},\ldots,e^{2\pi i \mu_N}\right)
\in Z(N) 
\ee
which belongs to the group center when $\mu$'s are given by \eq{condition3}. 

It follows from the general formulae of section II, that the effective action 
is a functional of the eigenvalues of $A_4$ in the {\em adjoint} representation.
If $A_4$ in the fundamental representation is diagonal and given by \eq{A4N},
the eigenvalues of the adjoint $(N^2\!-\!1)\times(N^2\!-\!1)$ matrix 
$A_{4\,{\rm adj}}^{ab}=if^{acb}A_4^c$ are
\be\label{adjointeigsN}
{\rm adjoint\; eigenvalues}=\pm 2\pi T(a_m-a_n)\equiv \pm 2\pi T\,\nu_{mn}.
\ee
There are $(N\!-\!1)$ zero eigenvalues whose number is the rank of the group,
and $N(N\!-\!1)$ pair-wise ($\pm$) non-zero eigenvalues. Under gauge transformation
\ur{discreteGT_N} the adjoint eigenvalues apparently shift by integers,
\be  
\nu_{mn}=(a_m-a_n)\to \nu_{mn}+(\mu_m-\mu_n)=\nu_{mn}+{\rm integers}
\ee
owing to \eq{condition2}. Therefore, the $Z(N)$ symmetry of the effective action
is manifest if it is periodic in adjoint eigenvalues $\nu_{mn}$ with period 1.

With $A_i$ set to zero, the calculation of the invariants (\ref{I1}-\ref{I3})
simplifies. Basically, they are sums of functionals of the
adjoint eigenvalues $\nu_{mn}$. The tree-level kinetic energy $E_i^aE_i^a/2=
\Tr(\partial_iA_4)^2=(2\pi T)^2\sum_{m=1}^N(\partial_ia_m)^2$ can be also
written as a sum over adjoint eigenvalues, using the identity
\be\label{identity}
\sum_{m=1}^N(\partial_ia_m)^2
=\frac{1}{N}\sum_{m<n}^N[\partial_i(a_m-a_n)]^2, 
\ee
which is satisfied  when ${\sum_1^Na_n=0}$ is taken into account. We thus obtain 
the following 1-loop effective action for the eigenvalues of the Polyakov
loop in a general $SU(N)$ theory:
\bea\label{eigenactionN}
S&=&-\sum_{m>n}^N\int\!d^3x\\
\nn
&&\times \left\{(\partial_i\nu_{mn})^2\frac{11}{12}T
\left[2\,\log\left(\frac{4\pi T}{\Lambda\,e^{\gamma_E}}\right)-H(\nu_{mn})\right]\right.\\ 
\nn 
&&\left.+\,\frac{(2\pi)^2T^3}{3}\nu_{mn}^2(1-\nu_{mn})^2\right\}, \quad \nu_{mn}=a_m-a_n,
\eea
where the function $H(\nu)$ is given by \eq{H1} and plotted in Fig.~2: it is periodic
with period 1. There are $N(N\!-\!1)/2$ similar terms in \eq{eigenactionN}, however
it should be kept in mind that there are only $N\!-\!1$ independent variables $a_m(x)$
through which the eigenvalues of the Polyakov line are expressed. At $N\!=\!3$ there
are three terms and the general result \ur{eigenactionN} comes to \eq{eigenaction}.
At $N\!=\!2$ there is only one term, and the result coincides with that of 
refs.~\cite{BGK-AP,DO2G}. 
The minima of the potential energy lie at integer values of $\nu_{mn}$ corresponding
to the Polyakov line belonging to the group center. The maxima 
correspond to the Polyakov line $L_{\rm max}={\rm diag}\left(e^{i\pi\frac{N-1}{N}},
e^{i\pi\frac{N-3}{N}},\ldots,e^{-i\pi\frac{N-1}{N}}\right)$ (and those which one gets from
this one by multiplying it by elements of the center), and it has the property that
$\Tr\,L_{\rm max}=0$. It is interesting that $H(\nu)$ is positive definite, so that the 
kinetic energy changes sign at certain values of $\nu$ depending on the temperature. 
It may signal an instability of the trivial (perturbative) vacua, as one lowers the temperature.    

\section{Conclusions}
We derived an effective 1-loop action for the Polyakov line in a pure
Yang--Mills theory in two settings: 1) for the Polyakov line as 
a unitary matrix rotating under spatial gauge transformations, 2) 
for the gauge-invariant eigenvalues of the Polyakov line. We do not assume
that the Polyakov line is close to the elements of the center, meaning that
we are collecting all powers of the $A_4$ field in the effective action 
while expanding in its (covariant) derivatives. 

In case  2 the  effective action is  an expansion in  ordinary spatial
derivatives  of  the  Polyakov  line eigenvalues,  and  is  explicitly
symmetric under discrete $Z(N)$ transformations of the eigenvalues. In
case 1 the  effective action is an expansion  in covariant derivatives
of $A_4$,  which necessarily include  the spatial components  $A_i$ of
the background field.  We expand the action to  include all invariants
of the type $E_i^2$ where $E_i$  is the electric field and of the type
$B_{i\parallel}^2$ where $B_i$ is  the magnetic field parallel, in the
SU(3) sense, to $A_4$. We have checked our results by reducing them to
the previously  studied $SU(2)$ case,  and by comparing them  with the
previously known expansion up to  quadratic terms in $A_4$. We stress,
however, that we collect all powers of $A_4$ in our effective action.

In case 1 our results are symmetric with respect to the group center 
only in part of the gauge invariants we compute but not in all of them. 
The reason is that the $Z(N)$ symmetry is actually a consequence of the 
invariance under fast time-dependent gauge transformations. 
Once $A_i$ fields are introduced to ensure gauge invariance of the effective action
under static transformations of the Polyakov line,  
the time-dependent transformations generate large time derivatives 
of $A_i$. Unless all powers in $\dot A_i$ are collected in the effective 
action, the $Z(N)$ symmetry is not manifest. Since we expand to the order 
of $E^2$ only (hence to the second order in $\dot A_i$) we cannot observe 
the $Z(N)$ symmetry. Our results in case 1 apply to static $(A_i,A_4)$ 
background fields and to the $SU(3)$ group only.

We hope that the results may be of some help to study correlation 
functions of the Polyakov line not too far from the transition
point where it experiences fluctuations that are
large in amplitude but presumably mainly long ranged.

\begin{appendix}
\section{Basics about $SU(3)$}
The generators of the gauge group $SU(3)$ are
\be
t^a = \frac{\lambda^a}{2} ,
\ee
where the matrices $\lambda^a$ are given by
\bea
\lambda^1 &=& \left(
\begin{array}{ccc}
0&1&0\\
1&0&0\\
0&0&0
\end{array}\right)\,,
\,\,\,
\lambda^2 = \left(
\begin{array}{ccc}
0&-i&0\\
i&0&0\\
0&0&0
\end{array}\right)\,,
\\\nonumber
\lambda^3 &=& \left(
\begin{array}{ccc}
1&0&0\\
0&-1&0\\
0&0&0
\end{array}\right)\,,
\,\,\,
\lambda^4 = \left(
\begin{array}{ccc}
0&0&1\\
0&0&0\\
1&0&0
\end{array}\right)\,,
\\\nonumber
\lambda^5 &=& \left(
\begin{array}{ccc}
0&0&-i\\
0&0&0\\
i&0&0
\end{array}\right)\,,
\,\,\,
\lambda^6 = \left(
\begin{array}{ccc}
0&0&0\\
0&0&1\\
0&1&0
\end{array}\right)\,,\\\nonumber
\lambda^7 &=& \left(
\begin{array}{ccc}
0&0&0\\
0&0&-i\\
0&i&0
\end{array}\right)\,,
\,\,\,
\lambda^8 = \frac{1}{\sqrt{3}}
\left(
\begin{array}{ccc}
1&0&0\\
0&1&0\\
0&0&-2
\end{array}\right).
\eea
The generators satisfy
\be
\left[t^a, t^b\right] = i f^{abc} t^c .
\ee
In addition to the totally anti-symmetric $f^{abc}$ there are totally
symmetric $d^{abc}$ structure constants which are defined according to
\be
\{t^a, t^b\} = \frac{1}{3}\delta^{ab} + d^{abc} t^c .
\ee
The non-vanishing values of $f^{abc}$ and $d^{abc}$ are summarized in Table
\ref{fd}.
\be\label{fd}
\begin{tabular}{cr|cr}\\\hline\hline
(a,b,c) & $f^{abc}$  & (a,b,c) & $d^{abc}$ \\\hline\hline
123 & 1 & 118 & $1/\sqrt{3}$\\  
147 & $1/2$ & 146 & $1/2$ \\  
156 & $-1/2$ & 157 & $1/2$ \\
246 & $1/2$ & 228 & $1/\sqrt{3}$ \\
257 & $1/2$ & 247 &$ -1/2$ \\
345 & $1/2$ & 256 & $1/2$ \\
367 & $-1/2$ & 338 & $1/\sqrt{3}$ \\
458 & $\sqrt{3}/2$ & 344 & $1/2$ \\ 
678 & $\sqrt{3}/2$ & 355 & $1/2$ \\
 & & 366 & $-1/2$ \\
 & & 377 & $-1/2$ \\
 & & 448 & $-1/(2\sqrt{3})$ \\ 
 & & 558 & $-1/(2\sqrt{3})$ \\ 
 & & 668 & $-1/(2\sqrt{3})$ \\ 
 & & 778 & $-1/(2\sqrt{3})$ \\ 
 & & 888 & $-1/(\sqrt{3})$ \\  \hline\hline
\end{tabular}
\ee
The $t^a$ matrices fulfill the following relations
\bea
t_{ij}^a t_{kl}^a &=& \half\left[\delta_{il}\delta_{jk} 
- \frac{1}{3}\delta_{ij}\delta_{kl}\right],\\
\Tr\, t^a &=& 0,\\
\Tr\left(t^a t^b\right) &=& \frac{\delta^{ab}}{2},\\
\Tr\left(t^a t^b t^c\right) &=& \frac{1}{4} \left(d^{abc} + i
  f^{abc}\right),\\
\Tr\left(t^a t^b t^a t^c\right) &=& - \frac{1}{12} \delta^{bc}.
\eea
The structure constants satisfy the Jacobi identities:
\bea\label{jac1}
f^{abe}f^{ecd} + f^{cbe}f^{aed} + f^{dbe}f^{ace} = 0,\\ \label{jac2}
f^{abe}d^{ecd} + f^{cbe}d^{aed} + f^{dbe}d^{ace} = 0.
\eea
Additionally
\be
f^{abe}f^{cde} =
\frac{2}{3}\left(\delta^{ac}\delta^{bd}-\delta^{ad}\delta^{bc}\right)
+\left(d^{ace}d^{bde} - d^{bce}d^{ade}\right).
\ee
By defining the $8\times 8$ matrices $F^a$ and $D^a$ such that
\bea
(F^a)^{bc} &=& - i f^{abc},\\ 
(D^a)^{bc} &=& d^{abc},
\eea
the Jacobi identities \urs{jac1}{jac2} take the form
\bea
\left[F^a, F^b\right] &=& i f^{abc} F^c,\\
\left[F^a, D^b\right] &=& i f^{abc} D^c.
\eea
And the fact that $f^{abb}=0$ and $d^{abb}=0$ implies
\be
\Tr\,F^a = 0\quad , \quad\Tr\,D^a = 0.
\ee
Some more useful relationships are:
\bea
f^{acd}f^{bcd} &=& 3 \delta^{ab},\\
F^a F^a &=& 3,\\
\Tr\,(F^a F^b)&=& 3 \delta^{ab},\\
f^{acd}d^{bcd} &=& 0,\\
F^a D^a &=& 0,\\
\Tr\,(F^a D^b)&=& 0,\\
d^{acd}d^{bcd} &=&\frac{5}{3}\delta^{ab},\\
D^a D^a &=&\frac{5}{3},\\
\Tr\,(D^a D^b)&=& \frac{5}{3} \delta^{ab},\\
\Tr\,(F^a F^b F^c)&=&i \frac{3}{2} f^{abc},\\
\Tr\,(D^a F^b F^c)&=& \frac{3}{2} d^{abc},\\
\Tr\,(D^a D^b F^c)&=& i\frac{5}{6} f^{abc},\\
\Tr\,(D^a D^b D^c)&=& -\frac{1}{2}d^{abc},\\
\Tr\,(F^a F^b F^a F^c)&=& \frac{9}{2}\delta^{bc}.
\eea
In particular for any matrix in the fundamental representation
\be
A = A^a t^a
\ee
one can construct the adjoint representation according to
\bea
A^{ab} = i f^{acb} A^c.
\eea

\end{appendix}




\begin{thebibliography}{999}
\bibitem{dimred}
T. Appelquist and R.D. Pisarski, \prd{23}{2305}{1981};
S. Nadkarni, \prd{27}{917}{1983}; \ibid{38}{3287}{1988};
N.P. Landsman, \npb{322}{498}{1989}.

\bibitem{P}
A.~M.~Polyakov, \plb{72}{477}{1978}.

\bibitem{L}
A.~D.~Linde, \plb{96}{289}{1980}.

\bibitem{GPY}
D.~J.~Gross, R.~D.~Pisarski and L.~G.~Yaffe, \rmp{53}{43}{1981}.

\bibitem{breakdown}
P.~Arnold and C.~Zhai, \prd{50}{7603}{1994}; \ibid{51}{1906}{1995};
E.~Braaten, \prl{74}{2164}{1995};
B.~Kastening and C.~Zhai, \prd{52}{7232}{1995};
E.~Braaten and A.~Nieto, \prd{53}{3421}{1996}.

\bibitem{S}
L.~Susskind, \prd{20}{2610}{1979}.

\bibitem{'tH}
G.'t~Hooft,  \npb{138}{1}{1978}; \ibid{153}{141}{1979}.

\bibitem{SY}
B.~Svetitsky and L.~G.~Yaffe, \npb{210}{423}{1982}.

\bibitem{Karsch}
O.~Kaczmarek, S.~Ejiri, F.~Karsch, E.~Laermann and F.~Zantow. {\em Talk given at Workshop on Finite Density QCD at Nara, Japan, 2003}. {\tt hep-lat/0312015}.

\bibitem{DHLOP}
A.~Dumitru, Y.~Hatta, J.~Lenaghan, K.~Orginos and R.~D.~Pisarski, Phys.~Rev.~D (to be published), {\tt hep-th/0311223}.

\bibitem{Weiss}
N.~Weiss, \prd{24}{475}{1981}; \ibid{25}{2667}{1982}.

\bibitem{2loop}
V.M. Belyaev and V.L. Eletsky, \jetp{50}{55}{1989}, \pzetf{50}{49}{1989};
K. Enqvist and K. Kajantie, \zpc{47}{291}{1990}.

\bibitem{DO2G}
D.~Diakonov and M.~Oswald, \prd{68}{025012}{2003}.

\bibitem{MO}
M.~Oswald, \appb{34}{5847}{2003}.

\bibitem{DO2F}
D.~Diakonov and M.~Oswald, Phys.~Rev.~D (to be published),{\tt hep-ph/0312126}.

\bibitem{DPY}
D.~Diakonov, V.~Yu.~Petrov and A.~V.~Yung, \plb{130}{385}{1983};
\yadf{39}{240}{1984}; \sjnp{39}{150}{1984}.

\bibitem{BGK-AP}
T. Bhattacharya, A. Gocksch, C. Korthals Altes and R.D. Pisarski, Nucl. Phys. B {\bf 383}
(1992) 497.

\bibitem{Schwinger}
J.~Schwinger, \pr{82}{664}{1951}.

\bibitem{coupling}
S.~Huang and M.~Lissia, \npb{438}{54}{1995};
K.~Kajantie, M.~Laine, K.~Rummukainen and M.~E.~Shaposhnikov,
\npb{458}{90}{1996}; \ibid{503}{357}{1997}.

\bibitem{chapman}
S.~Chapman, \prd{50}{5308}{1994}.

\bibitem{arriola}
E.~Megias, E.~Ruiz Arriola and
L.~L.~Salcedo, Phys.~Rev.~D (to be published), {\tt hep-ph/0312133}.

\bibitem{arriola2}
E.~Megias, E.~Ruiz Arriola and L.~L.~Salcedo, \plb{563}{173}{2003}.

\bibitem{Has}
A.~Hasenfratz and P.~Hasenfratz, \plb{93}{165}{1980}, \npb{193}{210}{1981}.


\end{thebibliography}
\end{document}